\documentclass[11pt,a4paper]{article}
\pdfoutput=1
\usepackage{jheppub}
\usepackage{textcomp}
\usepackage{amsmath}
\usepackage{mathtools}
\usepackage{mathrsfs}
\usepackage{filecontents}
\usepackage{braket,mleftright}
\usepackage{graphicx}
\usepackage{float}
\usepackage[position=t,singlelinecheck=off]{subfig}
\usepackage{caption}

\mleftright

\newcommand{\F}{ \textnormal{\begin{tiny}\textit{F}\end{tiny}}}

\title{Topology of Fermi Surfaces and Anomaly Inflows}%
\date{July 21 2016}

\affiliation[a]{Department of Mathematics, University of British Columbia, 1984 Mathematics Road, Vancouver, Canada V6T 1Z2}
\author[a]{Alejandro Adem}
\author[a]{Omar Antol\'in Camarena}
\author[b]{Gordon W. Semenoff}
\author[a]{Daniel Sheinbaum}

\affiliation[b]{Department of Physics and Astronomy, University of British Columbia, 6224 Agricultural Road, Vancouver, Canada V6T 1Z1}
\emailAdd{adem@math.ubc.ca}
\emailAdd{oantolin@math.ubc.ca}
\emailAdd{gordonws@phas.ubc.ca}
\emailAdd{dshein@math.ubc.ca}

\abstract{We derive a rigorous classification of topologically stable Fermi surfaces of non-interacting, discrete translation-invariant systems from electronic band theory, adiabatic evolution and their topological interpretations. For systems on an infinite crystal it is shown that there can only be topologically unstable Fermi surfaces. For systems on a half-space and with a gapped bulk, our derivation naturally yields a $\mathit{K}$-theory classification. Given the $d-1$-dimensional surface Brillouin zone $\mathrm{X}_{s}$ of a $d$-dimensional half-space, our result implies that different classes of globally stable Fermi surfaces belong in $\mathit{K^{-1}}\mathrm{(X_{s})}$ for systems with only discrete translation-invariance. This result has a chiral anomaly inflow interpretation, as it reduces to the spectral flow for $d = 2$. Through equivariant homotopy methods we extend these results for symmetry classes $AI,\,AII,\, C$ and $D$ and discuss their corresponding anomaly inflow interpretation.}

\begin{document}
\maketitle
\flushbottom
\section{Introduction}
Classification of locally stable Fermi surfaces and topological phases of matter using $\mathit{K}$-theory was first introduced in the pioneering work of Ho\v{r}ava \cite{Horava} and Kitaev \cite{Kitaev}. The present work generalizes to the infinite dimensional Hilbert space setting and global character the analysis of Ho\v{r}ava, though, through a rather different route. We do not restrict ourselves to analyzing the local character of perturbations around a chosen fixed-point of the Fermi surface. However, for the sake of completeness, we illustrate Ho\v{r}ava's analysis in this setting. The main novelty in our approach is the fact that, to our knowledge, we are the first to obtain directly a $\mathit{K}$-theoretic classification for global stability of Fermi surfaces without resorting to physically-unjustified mathematical procedures, as it has been done previously in the literature. The chiral anomaly connects to topology via the Atiyah-Patodi-Singer index theorem \cite{Atiyah-Spectral} and the notion of \textit{spectral flow}. This topological invariant comes out in our construction. Edge states of the Integer quantum Hall effect (IQHE) were already given a spectral flow interpretation by Fukui and collaborators \cite{Fukui-IQHE}. Here we put their result in the wider framework of $\mathit{K}$-theory and extend to other symmetry classes.

 Ryu et al. \cite{Ryu-Moore-Ludwig},\cite{Ryu-Zhang} and Witten \cite{Witten-FPI} among others, have developed a formalism for studying gapped bulk systems through classifying their gapless edge states and the \textit{bulk-boundary correspondence}, where the edge states have an anomaly inflow interpretation \cite{Callan-Harvey},\cite{Faddeev-Anomaly-Inflow}. This paradigm is physically similar to ours, however these arguments rely upon taking a low-energy limit of a lattice Hamiltonian to obtain an effective relativistic field theory \cite{Fradkin} such as Chern-Simons for the IQHE, with a correspondence between topological field theories and symmetry protected topological phases \cite{Witten-FPI},\cite{Wen-SPT}. The present work does not rely on such approximations, but we do have other strong assumptions such as discrete translation-invariance.

The results presented below show that systems on an infinite crystal cannot posses globally stable Fermi surfaces, up to homotopy. The proof presented here is, to our knowledge, new and does not rely on the usual arguments of fermion doubling \cite{Fermion-Doubling}. It is also quite simple from a mathematical point of view. We proceed to study systems on a half-space $\mathbb{R}^{d-1}\times [0,\infty)$. As discrete translation-invariance is broken by the surface $\mathbb{R}^{d-1}\times \{0\}$ only a reduced version of Bloch's theorem holds. Nevertheless this opens up the possibility of having surface states and so called bulk states become part of the continuous spectrum of our system. This, together with a gapped bulk condition will play the central role in the derivation of the $\mathit{K}$-theoretic classification of globally stable Fermi surfaces for systems on a half-space.

Issues of \textit{aperiodicity, disorder} and \textit{unitary-antiunitary} ambiguities in representations of symmetry operators are considered in a more general framework in \cite{Freed-Moore},\cite{Thiang1} and will not be dealt with in this work. However, neither of \cite{Freed-Moore},\cite{Thiang1} attempt to classify Fermi surfaces. We work in the so-called \textit{first quantization} formalism where we construct the ground state of our systems employing the \textit{single-particle} picture and filling the lowest single-particle states using Pauli's exclusion principle. 

\section{Momentum-space classification with only discrete translation-symmetry}
We proceed to derive a classification for both globally topologically stable Fermi surfaces for systems on an infinite crystal in subsection \ref{Sub:IC} and half-spaces $\mathbb{R}^{d-1}\times [0,\infty)$ in subsection \ref{Sub:HS}. To put our classification in the adequate context, we include in subection \ref{Horava-Local} a simplified version of Ho\v{r}ava's classification for locally stable Fermi surfaces in the single-particle setting, as well as its limitations for modeling globally stable surfaces. By a globally topologically \textit{stable} Fermi surface we mean that its associated gapless system cannot become gapped (have an empty Fermi surface) as long as certain conditions, carefully detailed below, are satisfied. Otherwise the Fermi surface is referred to as topologically globally \textit{unstable}.

 All systems considered here are sometimes treated in the literature in the \textit{second quantization} formalism, constructed in a fermionic \textit{Fock} space of particle creation/annihilation operators. As stated previously, we work under the single-particle picture and we have adapted Ho\v{r}ava's local classification into first quantization form, which allows us to encompass the different systems with the same formalism and also permits a simpler interpretation of the operator theoretic aspects of the problem, as in \cite{Freed-Moore},\cite{Thiang1}.
\subsection{Infinite crystal}\label{Sub:IC}
Let $\mathrm{X}$ be the Brillouin zone of a non-interacting fermionic system on an infinite crystal, i.e. $\mathbb{R}^d$. For each crystal momentum $\vec{k} \in \mathrm{X}$ there is a Bloch Hamiltonian operator $\mathcal{H}(\vec{k})$ acting on a complex separable Hilbert space $\mathscr{H}$. $\mathcal{H}(\vec{k})$ is self-adjoint and there exists a complete orthonormal basis $\{\ket{ \Phi_{n}(\vec{k})} \}_{n=1}^{\infty}\,$ of $\mathscr{H}$, such that
\begin{eqnarray}
\mathcal{H}(\vec{k})\ket{ \Phi_{n}(\vec{k})} &=&\varepsilon_{n}(\vec{k})\ket{ \Phi_{n}(\vec{k})}\,,\\
\mathcal{H}(\vec{k}) &=&  (i\nabla - \vec{k})^2 + V(\vec{r})\,,\label{Infinite-Crystal-Bloch}\\
V(\vec{r}) &=& V(\vec{r}+\vec{R})\,
\end{eqnarray}
where the $\vec{k}$-dependent eigenvalues $\varepsilon_{n}(\vec{k})$ are known as energy bands and $\vec{R}$ is a lattice vector. 

 The Fermi surface \cite{Ashcroft-Mermin} of a non-interacting fermionic system corresponds to the set of points in the Brillouin zone for which one or more energy bands are equal to the Fermi energy $\varepsilon_{\F}$
\begin{eqnarray}
\left \{ \vec{k}\in \mathrm{X} \,|\,(\mathcal{H}(\vec{k}) - \varepsilon_{\F} I)\ket{ \Phi_{n}(\vec{k})} = 0\,, \text{ for some }n \right \} \label{FS-def}
\end{eqnarray} 
 Hence the operator of interest is $(\mathcal{H}(\vec{k}) - \varepsilon_{\F} I)$ and we wish to study its kernel, which has a physical interpretation. We have the freedom to choose our energy scale such that $\varepsilon_{\F} = 0$ and we only write $\mathcal{H}(\vec{k})$ instead. If we restrict ourselves to the $n$-th energy band $\varepsilon_{n}(\vec{k})$, the corresponding surface embedded in the Brillouin zone is called the $n$-th \textit{branch} of the Fermi surface, the Fermi surface being the union of all such branches. We assume that none of our Bloch Hamiltonians are infinitely degenerate at the Fermi energy, thus
\begin{equation}
\mathit{dim\,Ker}\, \mathcal{H}(\vec{k}) < \infty\,\,\,\,\forall \,\vec{k} \in \mathrm{X}\,.
\end{equation}
The work of Atiyah and Singer \cite{Atiyah-Skew} is central to our classification. To enable the use of the machinery developed in \cite{Atiyah-Skew} we must impose that our Hamiltonian operators be \textit{bounded}. This is a standard trick employed when analyzing topological properties of Hamiltonian and Dirac operators, which is usually done by employing the topology induced by the \textit{Riesz} norm, defined by imposing that the transformation
 \begin{equation}
 \mathcal{H}(\vec{k}) \mapsto \mathcal{H}(\vec{k}) (I +\mathcal{H}(\vec{k})^{\dagger}\mathcal{H}(\vec{k}) )^{-\frac{1}{2}}
\end{equation}
be an \textit{isometry}, making unbounded operators bounded \cite{Lesch-Riesz}. Thus, we could avoid the unnatural bounded condition by working in the so called Riesz topology, but for the sake of simplicity we shall stick with the operator norm topology for the time being. We give the details of the equivalence of both treatments in Appendix \ref{Riesz-Topology}.
\pagebreak
 We list the assumed properties of our Bloch Hamiltonian in mathematical terms:
\begin{itemize}
\item $\mathcal{H}(\vec{k})$ is \textit{bounded},
\item $\mathit{dim\,Ker}\, \mathcal{H}(\vec{k}) < \infty\,$,
\item $\mathcal{H}(\vec{k})$ is \textit{self-adjoint}.
\end{itemize}
\nopagebreak
Operators satisfying these conditions are known in the mathematical literature as bounded \textit{self-adjoint} \textit{Fredholm} operators \cite{Atiyah-Skew}, where the set of bounded Fredholm operators \cite{Atiyah-Fredholm} are bounded operators in $\mathscr{H}$ with finite kernel and cokernel. We denote the space of all complex bounded Fredholm operators $\mathcal{F}(\mathscr{H})$ and bounded self-adjoint Fredholm operators as $\mathcal{F}^{sa}(\mathscr{H})$. 
\subsubsection{Path-components of the space of bounded self-adjoint Fredholm operators}
Let us endow $\mathcal{F}(\mathscr{H})$ with the \textit{norm} topology, the most natural topology for bounded Fredholm operators (see Appendix \ref{Riesz-Topology}).
With this choice of topology $\mathcal{F}^{sa}(\mathscr{H})$ has 3 disjoint path-components, operators which are \textit{essentially positive} denoted $\mathcal{F}^{sa}_{+}(\mathscr{H})$, \textit{essentially negative} $\mathcal{F}^{sa}_{-}(\mathscr{H})$ and the remaining path-component $\mathcal{F}^{sa}_{*}(\mathscr{H})$. Care must be given to the subtleties of what it means to be essentially positive/negative. A self-adjoint operator $\mathcal{B}$ is positive if
\begin{equation}
\bra{\phi}\mathcal{B}\ket{\phi} \geq 0 \,\,\forall \ket{\phi} \in \mathscr{H}\,,
\end{equation}
and essentially positive if $\exists\, \mathscr{V}\subset \mathscr{H}$ with $codim \,\mathscr{V} < \infty$ such that $\mathcal{B}$ is positive on $\mathscr{V}$ and $\mathcal{B}\mathscr{V} = \mathscr{V}$ \cite{Atiyah-Skew},\cite{Boos-Essential}. The definition of essentially negative is analogous. Depending on which path-component our Hamiltonian belongs to we will obtain very different results.

\subsection{Adiabatic evolution of systems on an infinite crystal}
$\mathcal{H}(\vec{k})$ is a continuous function of $\vec{k}$, so its image lies in one of the 3 disjoint path-components of $\mathcal{F}^{sa}(\mathscr{H})$. With this in mind, we wish to put in the same equivalence relation all systems which can be \textit{adiabatically evolved} \cite{Avron-Adiabatic} into one another. As usually done in the literature, we employ a \textit{homotopy} model of adiabatic evolution, i.e. $\forall\,\, \mathcal{H}_{0},\mathcal{H}_{1}: \mathrm{X}\rightarrow\mathcal{F}^{sa}(\mathscr{H})$ we have:
\begin{eqnarray}
\mathcal{H}_{0}\sim\mathcal{H}_{1}\Longleftrightarrow \exists \, g: \mathrm{X}\times I\rightarrow \mathcal{F}^{sa}(\mathscr{H}) \textit{ continuous,  } \nonumber \\
g(\vec{k},0) =\mathcal{H}_{0}(\vec{k}),\,\,\, g(\vec{k},1) = \mathcal{H}_{1}(\vec{k}) \,\,\forall \,\,\vec{k}\in \mathrm{X}\,.\label{Adiabatic-Homotopy}
\end{eqnarray}
There is a rigorous notion of adiabatic evolution in condensed matter systems given by Avron et al. \cite{Avron-Adiabatic} with applications to the  IQHE, and it has been generalized to systems without a gap condition \cite{Avron-Gapless}. We emphasize that eq.(\ref{Adiabatic-Homotopy}) is a sufficiently good approximation to \cite{Avron-Adiabatic} only when we restrict the space of allowed Hamiltonians and is in general not true for arbitrary ones. We also remark that for metals the notion of adiabatic evolution should be understood in the context of Landau's Fermi liquid theory \cite{Schulz-Fermi-Liquid} as in \cite{Horava},\cite{Ryu-Fermi},\cite{Chiu-Classification}.
	
	Let us denote by $\sigma(\mathcal{H}(\vec{k}))$ the \textit{spectrum} of $\mathcal{H}(\vec{k})$. On an infinite crystal Bloch's theorem induces the following periodicity condition
\begin{equation}
\phi_n(\vec{k},\vec{r}+\vec{R}) = \phi_n(\vec{k},\vec{r}),
\end{equation}
were  again $\vec{R}$ is an element of the direct lattice that represents the crystal. Let us remember that the labeling of solutions by the crystal momentum $\vec{k}$ is due to the unitarity of the translation operator $T_{\vec{R}}$ and Bloch's theorem \cite{Feldman-Spectrum} and since $\vec{k}\in \mathrm{X}$, $\vec{k}$ has \textit{real} components. Furthermore, if $V(\vec{r}) = V(\vec{r}+\vec{R}) $, is also bounded, then $\mathcal{H}(\vec{k})$ has \textit{compact resolvent}, implying that the \textit{continuous} spectrum of $\mathcal{H}(\vec{k})$,  $\sigma_{c}(\mathcal{H}(\vec{k}))$ is empty. Thus the spectrum of $\mathcal{H}(\vec{k})$ coincides with its eigenvalues. Since $\mathcal{H}(\vec{k})$ is bounded from below, it has a finite number of valence bands (negative eigenvalues for each $\vec{k}$) \cite{Feldman-Spectrum}, i.e. $\mathcal{H}(\vec{k}) \in \mathcal{F}^{sa}_{+}(\mathscr{H})$ after applying a Riesz transformation (Appendix \ref{Riesz-Topology}). Thus, all types of Fermi surfaces for systems on an infinite crystal, separated by equivalence relation (\ref{Adiabatic-Homotopy}) are given by the set of \textit{homotopy classes of maps} $[\mathrm{X}, \mathcal{F}^{sa}_{+}(\mathscr{H})]$.
However, Atiyah and Singer proved in \cite{Atiyah-Skew} that $\mathcal{F}^{sa}_{+}(\mathscr{H})$ is homotopically equivalent to a point via the simple homotopy 
\begin{equation}
\mathcal{H}_{t}(\vec{k}) = (1-t)\mathcal{H}(\vec{k}) + t\mathcal{H}_{0}(\vec{k})\,,
\end{equation}
where $\mathcal{H}_0(\vec{k})$ is any gapped bounded Bloch Hamiltonian $\forall\, \vec{k} \in \mathrm{X}$. This implies
\begin{eqnarray}
[\mathrm{X}, \mathcal{F}^{sa}_{+}(\mathscr{H})] &=& *\,,\label{FR-trivial}
\end{eqnarray}
$*$ denoting a single, i.e. trivial class. Hence, all systems as described above can be connected to a gapped system through adiabatic evolution. This leads us to the following conclusion: \textit{all systems in symmetry classes $A,\,AI,$ and $AII$, on an infinite crystal, only have globally topologically unstable Fermi surfaces, up to adiabatic evolution}. Class $A$ refers to systems whose only symmetry is discrete translation-invariance in the Altland-Zirnbauer classification \cite{Altland-Zirnbauer}. The validity for the remaining symmetry classes $AI$ and $AII$ will be discussed in section \ref{Sec:Symmetries}. 

This result was expected by arguments similar to those of fermion doubling on lattice gauge theories \cite{Fermion-Doubling}, by the Nielsen-Ninomiya theorem, where the Hamiltonians under consideration are matrices. For a recent discussion of the role of the Nielsen-Ninomiya theorem in condensed matter see \cite{Witten-Lectures}. Note that though our result yields sufficient conditions to guarantee the topological instability of the Fermi surface of a system on an infinite crystal, it does not have anything to say about necessary conditions. In the argument presented above we used the fact that for the infinite crystal, the spectrum of $\mathcal{H}(\vec{k})$ is discrete and bounded from below, which guarantees that $\mathcal{H}(\vec{k}) \in \mathcal{F}^{sa}_{+}(\mathscr{H})$. These are not necessary conditions as there can also be unbounded Dirac Hamiltonians, which do not belong to $\mathcal{F}^{sa}_{+}(\mathscr{H})$ but still correspond to a trivial class and have a globally topologically unstable Fermi surface. We also remark that the topology of the Brillouin zone was irrelevant for this result since what produces the triviality is that $\mathcal{F}^{sa}_{+}(\mathscr{H})$ is \textit{contractible} \cite{Hatcher-Alg-Top}.

 Finally, even though the physical result was expected, the method employed here is new to our knowledge, and perhaps the simplest and most straightforward method to prove such a result. It does not depend on the dimension of the system and does not depend on taking a low-energy limit of a lattice Hamiltonian \cite{Fradkin}.

\subsection{Local stability of Fermi surfaces}\label{Horava-Local}
Ho\v{r}ava \cite{Horava}  originally developed a classification of locally stable Fermi surfaces, that is, pieces of the Fermi surfaces which are robust to small perturbations near a given $\vec{k}_{0}$ on the Fermi surface. Later on Matsuura et al. \cite{Ryu-Fermi} and Chiu et al. \cite{Chiu-Classification} extended Ho\v{r}ava's work to include more symmetry classes. Both \cite{Ryu-Fermi},\cite{Chiu-Classification} clarify that they restrict to a neighborhood around a point because of the similar arguments to the above ones concerning fermion doubling. From our construction it would seem that restricting to a subspace of $\mathrm{X}$ does not yield a different result since (\ref{FR-trivial}) is true for any choice of space $\mathrm{X}$. What we must do is also restrict the set of allowed homotopies. Let us denote one of the path-components of the Fermi surfaces by $\Sigma$ and let $dim\,\Sigma = d-p-1$. One looks at the $p+1$-transverse directions to $\Sigma$ in $\mathrm{X}$ around a chosen point $\vec{k}_{0} \in \Sigma$. Ho\v{r}ava's resolution, for which we present a simplified version, adapted to the infinite dimensional formalism, is to consider the boundary of a ball $\partial B_{r_0}(\vec{k}_{0}) \subset \mathrm{X}$, which does not intersect $\Sigma$. The boundary is a topological sphere $\mathrm{S}^{p}$, whose radius $r_0\equiv r_0(\vec{k_{0}})$ depends on the choice of $\vec{k}_{0} \in \Sigma$. As long as we impose that the adiabatic evolution of the system does not generate a zero eigenvalue for any $\mathcal{H}(\vec{k})$, with $\vec{k} \in \mathrm{S}^{p}$, we have restricted homotopy classes of maps
\begin{eqnarray}
\label{Freed-adaptedto-Horava}
[\mathrm{S}^{p}, \mathcal{TP}(\mathscr{H})] &\approx & \mathit{Vect}_{\mathbb{C}}(\mathrm{S}^{p}).
\end{eqnarray}
where $\mathcal{TP}(\mathscr{H}) \subset \mathcal{F}^{sa}_{+}(\mathscr{H})$ denotes the space of \textit{invertible} Hamiltonians \cite{Freed-Moore} with empty continuous spectrum, also called \textit{gapped}, i.e. 
\begin{eqnarray}
dim\,Ker\, \mathcal{H}(\vec{k}) &=& 0\,\,\,\,\forall\,\mathcal{H}(\vec{k}) \in \mathcal{TP}(\mathscr{H})\,, \\
\sigma_c(\mathcal{H}(\vec{k})) &=& \phi\,.
\end{eqnarray}

These operators were widely studied in \cite{Freed-Moore},\cite{Thiang1},\cite{D-G-AI},\cite{D-G-AII}. $\mathit{Vect}_{\mathbb{C}}(\mathrm{S}^{p})$ is set of isomorphism classes of complex \textit{vector bundles} with base space $\mathrm{S}^{p}$ \cite{Hatcher-K-theory}. Equation (\ref{Freed-adaptedto-Horava}) can be derived by repeating a simplified version of the analysis in \cite{Freed-Moore}, section $10$. The arguments presented there follow through, except that in our case, we must restrict the parameter space for our families of Bloch Hamiltonians to $\partial B_{r_0}(\vec{k}_{0}) \approx \mathrm{S}^{p}$ instead of the whole Brillouin zone $\mathrm{X}$. If we are willing to lose information on the homotopy classes but gain computability via a \textit{stabilization} (see \cite{Freed-Moore},\cite{Hatcher-K-theory} for details, or \cite{D-G-AI},\cite{D-G-AII} for an approach without stabilization) we obtain the reduced $\mathit{K}$-theory group 
\begin{equation}
\mathit{\tilde{K}}(\mathrm{S}^{p}) =   
\begin{cases*} \mathbb{Z} & $p$ even\\
 0 & otherwise. \end{cases*}
\end{equation}

Here $\mathit{\tilde{K}}(\mathrm{S}^{p})$ denotes the \textit{reduced} Grothendieck completion of $\mathit{Vect}_{\mathbb{C}}(\mathrm{S}^{p})$, the semi-group of isomorphim classes of complex vector bundles with base space $\mathrm{S}^{p}$ \cite{Hatcher-K-theory}. 
\subsubsection{Limitations of local stability}
In order to avoid our first conclusion about the topological triviality of Fermi surfaces on an infinite crystal, we had to restrict the set of allowed homotopies in the analysis presented above. There is no natural nor universal procedure to make such a restriction for all $\vec{k}\in \mathrm{X}$ as there is no analogous quantity that would play the role of the gap in the bulk spectrum for topological phases for the whole Brillouin zone. Thus, we would have to look at the particularities of each chosen point $\vec{k}_{0}\in\Sigma$ to define an adequate $\mathit{S}^{p}$. In particular this construction is not well suited for modeling globally stable gapless edge modes observed at the boundaries of topological phases of matter. That being said, because of the role of $\mathcal{TP}(\mathscr{H})$ in Ho\v{r}ava's analysis, under the right codimension $d-p-1$, the local analysis \textit{emulates} part of the global analysis of gapped topological phases (see \cite{Ryu-Fermi},\cite{Chiu-Classification},\cite{Zhao-defects} for a discussions and connections) but it strongly fails to account for all the involution fixed-points induced by symmetries \cite{Chiu-Classification}, which give rise to distinct topological phases \cite{Freed-Moore}.
 
\subsection{Systems on a half-space}\label{Sub:HS}
Let us now consider systems on the half-space $\mathbb{R}^{d-1}\times[0,\infty)$, i.e. systems with a single boundary. Typically the boundary is a domain wall between two different systems (or vacuums). Let us denote the directions parallel to the the boundary by $\vec{r}_{\parallel}$. Since we no longer have the periodicity of the potential in the direction perpendicular to the boundary (denoted as $r_{\perp}$), we can only apply a partial Bloch theorem, in the directions parallel to the boundary. We set the boundary at $r_{\perp} = 0$. Our wave functions obey
\begin{eqnarray}
\psi_{\alpha}(\vec{k}_{\parallel},\vec{r}_{\parallel},r_{\perp}) &=& e^{i\vec{k}_{\parallel}\centerdot\vec{r}_{\parallel}}\phi_{\alpha}(\vec{k}_{\parallel},\vec{r}_{\parallel}, r_{\perp})\,,\\
T_{\vec{R}_{\parallel}} \ket{\phi_{\alpha}(\vec{k}_{\parallel},r_{\perp})} &=&  \ket{\phi_{\alpha}(\vec{k}_{\parallel}, r_{\perp})}\,.
\end{eqnarray}

Where $\vec{R}_{\parallel}$ is a surface lattice vector and $T_{\vec{R}_{\parallel}}$ is the associated discrete translation operator. Let us remark that, similarly to the infinite crystal case, the labeling of solutions by $\vec{k}_{\parallel}$ comes from the unitarity of $T_{\vec{R}_{\parallel}}$ and hence we can restrict $\vec{k}_{\parallel}$ to be in $\mathrm{X}_{s}$, where $\mathrm{X}_{s}$ is the Brillouin zone associated to the surface reciprocal lattice. There are different boundary conditions that can be imposed at $r_{\perp} = 0$, depending on the system under consideration. In general one must match a solution with the same functional form as an infinite crystal solution (i.e with no boundary) to solutions of the free Schr\"odinger equation (or more generally to the solutions of a Schr\"odinger equation in a different crystal), which decays exponentially as $r_{\perp}$ becomes positive and we leave the crystal \cite{Ashcroft-Mermin},\cite{Gross-Surface},\cite{Davison-Surface}. The infinite crystal-like piece of $\psi_{\alpha}(\vec{k}_{\parallel})$ has an $e^{ik_{\perp}r_{\perp}}$ factor, where $k_{\perp}$ originally represented the component of the crystal momentum corresponding to $r_{\perp}$. If we had an infinite crystal, $k_{\perp}$ would be forced by unitarity of $T_{\vec{R}}$ to be real. Nonetheless, since the boundary breaks the discrete translational symmetry in the $r_{\perp}$-direction, $k_{\perp}$ is no longer a good quantum number and it can take complex values as well (unlike the components of $\vec{k}_{\parallel}$, which must be real by construction) \cite{Ashcroft-Mermin},\cite{Gross-Surface},\cite{Davison-Surface}. These complex values of $k_{\perp}$ allow for the existence of new kinds of solutions described below. We thus transform the single-particle Schr\"odinger equation into one for each $\vec{k}_{\parallel}$ of the form
\begin{eqnarray}
\mathcal{H}(\vec{k}_{\parallel})\ket{\phi_{\alpha}(\vec{k}_{\parallel},r_{\perp})} = \varepsilon_{\alpha}(\vec{k}_{\parallel})\ket{\phi_{\alpha}(\vec{k}_{\parallel}, r_{\perp})}\,,\\
\mathcal{H}(\vec{k}_{\parallel}) = -\partial^{2}_{r_{\perp}} + (i\nabla_{\parallel}- \vec{k}_{\parallel})^2 + V(\vec{r}_{\parallel},r_{\perp})\,,\label{Half-space-Bloch}\\
V(\vec{r}_{\parallel},r _{\perp}) = V(\vec{r}_{\parallel}+\vec{R}_{\parallel}, r_{\perp})\,.
\end{eqnarray}
The index $\alpha \equiv \alpha (k_{\perp},m)$ behaves similarly to the \textit{principal quantum number} $n$ of the hydrogen atom, i.e. $\varepsilon_{\alpha}(\vec{k}_{\parallel})$ can either be part of the continuous spectrum of $\mathcal{H}(\vec{k}_{\parallel})$, $\sigma_{c}(\mathcal{H}(\vec{k}_{\parallel}))$ or part of the discrete spectrum $\sigma_{d}(\mathcal{H}(\vec{k}_{\parallel}))$.

\begin{figure}[]
%\captionsetup[subfigure]{labelformat=parens}
  	\centering
  	\captionsetup{justification=raggedright}
  	%\vspace{1cm}
  \subfloat[ ]{\label{fig:Bulk-State}\includegraphics[width=0.3\textwidth]{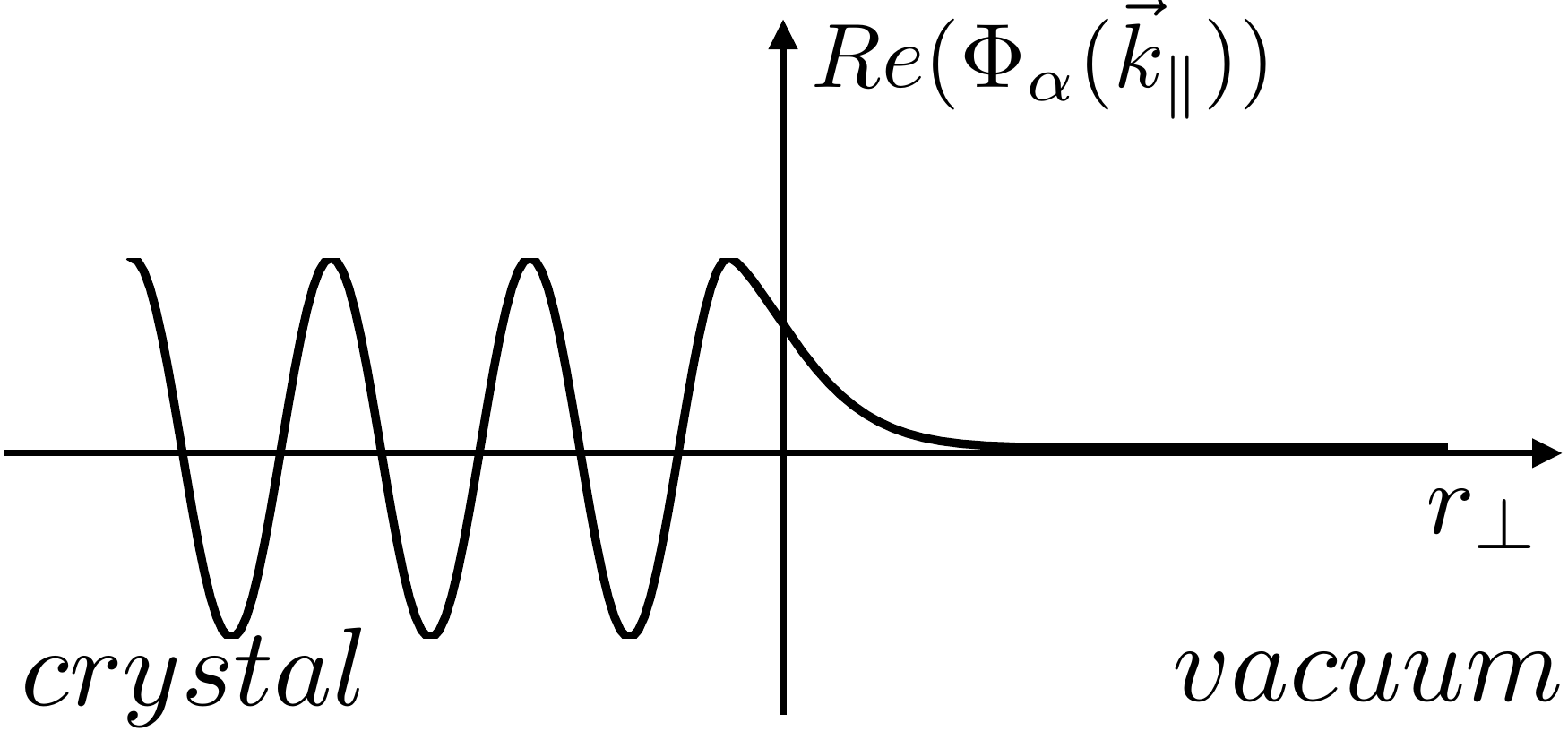}}\\
  \subfloat[ ]{\label{fig:Surface-State}\includegraphics[width=0.3\textwidth]{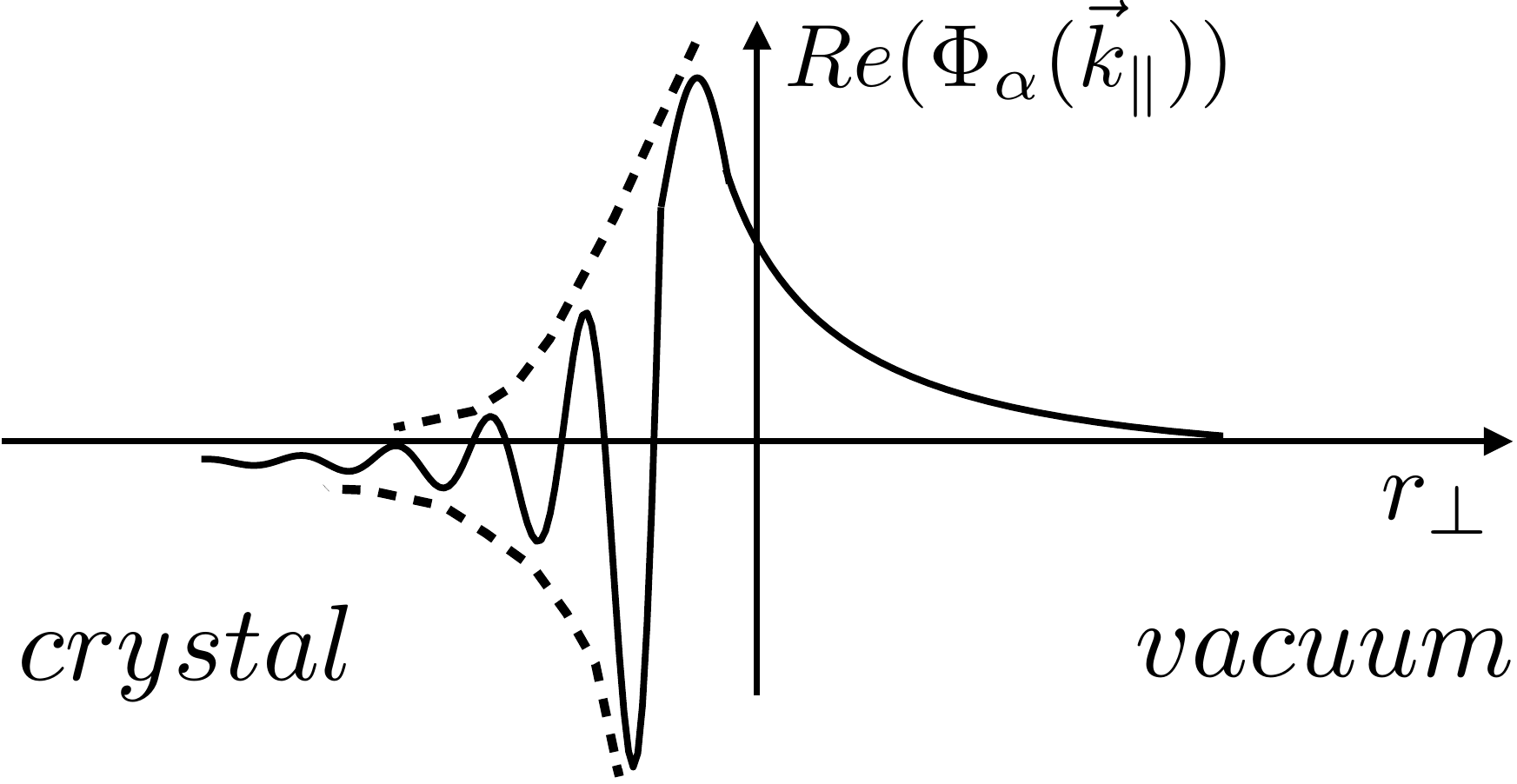}}\\
   \subfloat[ ]{\label{fig:Resonance-State}\includegraphics[width=0.3\textwidth]{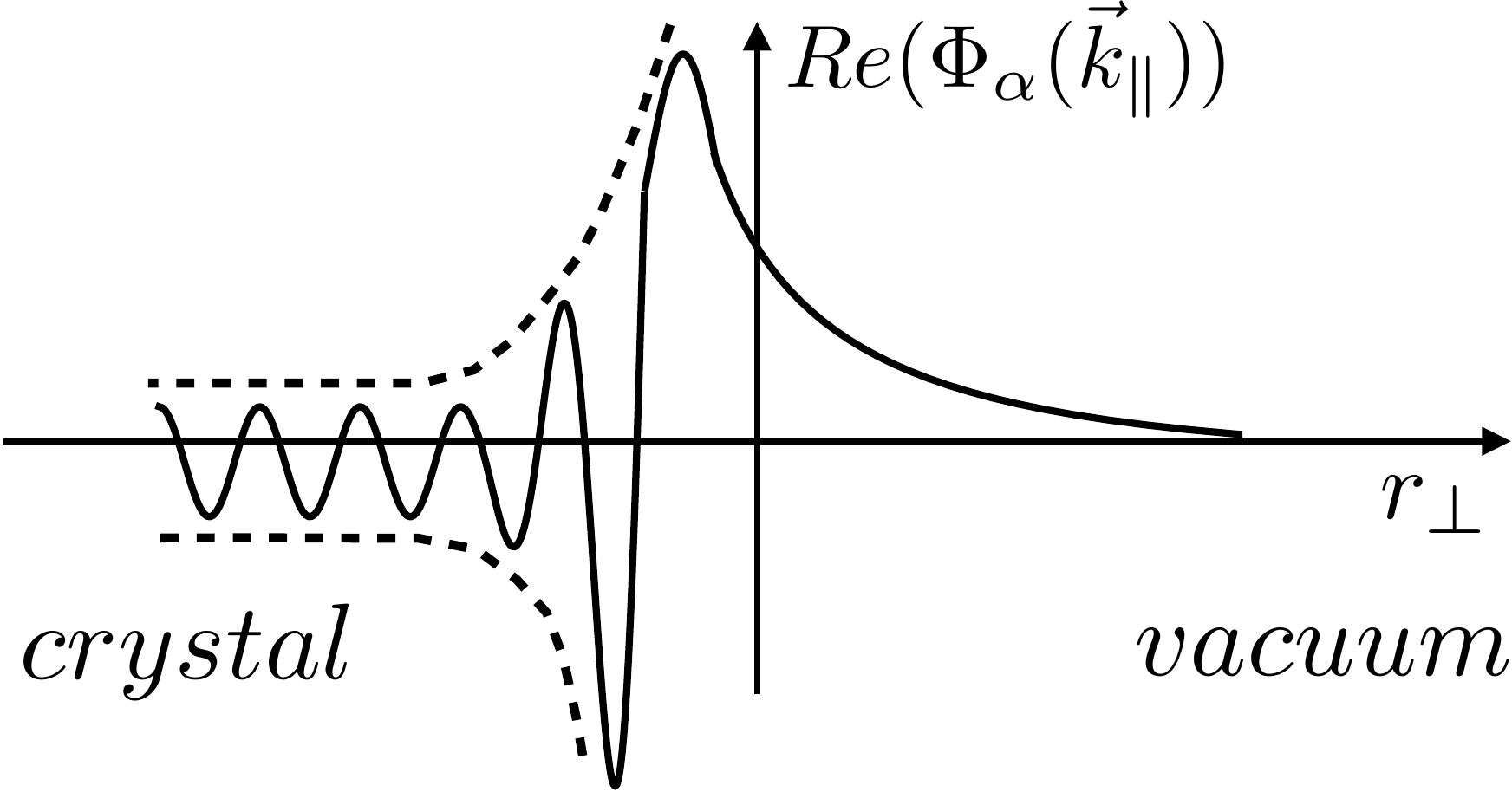}}
	 	 \caption[]{ Real part of the wave function of the distinct types of states on a half-space as a function of $r_{\perp}$. All solutions decay exponentially outside the crystal. \ref{fig:Bulk-State} is a bulk state which is periodic inside the crystal. \ref{fig:Surface-State} Surface state, decays very rapidly away from the boundary. \ref{fig:Resonance-State} Resonance state, the decay eventually stops and the solution is again periodic. Dashed lines represent the \textit{envelope} of the solutions. }
	 	 \label{FigureI}
\end{figure}

There are 3 types of solutions to the associated PDE problem, represented in figure \ref{FigureI}. Solutions with real $k_{\perp}$ ($Im(k_{\perp}) = 0$) are called \textit{bulk states} and correspond to truncating at the boundary the periodic solutions of the original infinite crystal. These can have any real value of $k_{\perp}$ and thus, they are part of $\sigma_{c}(\mathcal{H}(\vec{k}_{\parallel}))$. Note that though the PDE problem has a solution, the eigenvalue equation has no solution in $\mathscr{H}$ for $\varepsilon_{\alpha}(\vec{k}_{\parallel}) \in \sigma_{c}(\mathcal{H}(\vec{k}_{\parallel}))$, similar to the scattering states of the hydrogen atom \cite{Takhtajan-QM}. Solutions with $Im(k_{\perp}) \neq 0$ not only decay exponentially outside the crystal but also have an exponentially decaying envelope in the $r_{\perp}$-direction, towards the crystal 
\begin{eqnarray}
 \phi_{\alpha}(\vec{k}_{\parallel}, \vec{r}_{\parallel},r_{\perp}) \approx e^{-|Im(k_{\perp}) r_{\perp}|}u_{\alpha}(\vec{k}_{\parallel},\vec{r}_{\parallel}) 
\end{eqnarray}

where $u_{\alpha}(\vec{k}_{\parallel},\vec{r}_{\parallel})$ is a bounded and periodic function of $\vec{r}_{\parallel}$. These solutions are called \textit{surface states} \cite{Ashcroft-Mermin},\cite{Gross-Surface},\cite{Davison-Surface}. Matching such solutions to vacuum solutions at the boundary forces $k_{\perp}$ to come in discrete values, implying that surface states correspond to $\varepsilon_{\alpha}(\vec{k}_{\parallel}) \in\sigma_{d}(\mathcal{H}(\vec{k}_{\parallel}))$ \cite{Ashcroft-Mermin},\cite{Davison-Surface}. $m \in \mathbb{Z}$ is the number that arises from having Bloch's theorem in $d-1$ dimensions. Note that surface states do belong in $\mathscr{H}$ as they are \textit{normalizable}. Finally, there can be solutions which join a surface state with a bulk state, called \textit{surface resonance states} and also form part of $\sigma_{c}(\mathcal{H}(\vec{k}_{\parallel}))$. This joining is a limiting process in which a bulk/surface state becomes a surface/bulk state at some $\vec{k}_{\parallel} \in \mathrm{X}_{s}$, i.e. $u_{\alpha}(\vec{k}_{\parallel},\vec{r}_{\parallel})$ continues to be periodic but stops being bounded by $e^{-|Im(k_{\perp})r_{\perp}|}$. 

\subsection{Adiabatic evolution of systems on half-space}
If we impose that the bulk be \textit{gapped}, the Fermi energy $\varepsilon_{\F}$ must lie between bulk states, i.e. $\varepsilon_{\F}  \notin \sigma_{c}(\mathcal{H}(\vec{k}_{\parallel}))$. Setting $\varepsilon_{\F} = 0$, this implies that $\sigma_{c}(\mathcal{H}(\vec{k}_{\parallel}))\cap \mathbb{R}^{-} \neq \phi$, corresponding to the bulk valence band nearest to the Fermi energy and so $\mathcal{H}(\vec{k}_{\parallel}) \notin \mathcal{F}^{sa}_{+}(\mathscr{H})$. Similarly $\sigma_{c}(\mathcal{H}(\vec{k}_{\parallel}))\cap \mathbb{R}^{+} \neq \phi$, corresponding to the bulk conduction band nearest to the Fermi energy, which implies that $\mathcal{H}(\vec{k}_{\parallel}) \notin \mathcal{F}^{sa}_{-}(\mathscr{H})$. Thus, $\mathcal{H}(\vec{k}_{\parallel}) \in \mathcal{F}^{sa}_{*}(\mathscr{H})$ as long as the bulk remains gapped. Globally stable Fermi surfaces of systems on a half-space are thus classified by the set of homotopy classes $[\mathrm{X}_{s}, \mathcal{F}^{sa}_{*}(\mathscr{H})]$.

At this point let us remark that adiabatic evolution for systems with metallic surfaces refers to a bulk adiabatic evolution, i.e. bulk states below $\varepsilon_{\F}$ cannot tunnel directly to bulk states above the Fermi energy under adiabatic evolution, as defined in eq.(\ref{Adiabatic-Homotopy}). Nonetheless bulk/surfaces states are allowed to tunnel to surface/bulk states through surface resonances.

The set of skew-adjoint Fredholm operators $\hat{\mathcal{F}}(\mathscr{H})\equiv i\mathcal{F}^{sa}(\mathscr{H})$ is trivially \textit{homeomorphic} to $\mathcal{F}^{sa}(\mathscr{H})$ in the norm topology, and so are $\mathcal{F}^{sa}_{*}(\mathscr{H})$ and $\hat{\mathcal{F}}_{*}(\mathscr{H})$, the corresponding non-trivial component of $\hat{\mathcal{F}}(\mathscr{H})$.
   Atiyah and Singer proved in \cite{Atiyah-Skew} that $\hat{\mathcal{F}}_{*}(\mathscr{H}) \simeq \Omega\mathcal{F}(\mathscr{H})$, the \textit{loop} space of the set of Fredholm operators $\mathcal{F}(\mathscr{H})$. Combining this with Atiyah's proof \cite{Atiyah-Fredholm} that $\mathcal{F}(\mathscr{H})$ is a \textit{classifying} space for complex $\mathit{K}$-theory and the \textit{suspension isomorphism} \cite{Hatcher-Alg-Top} we get
\begin{eqnarray}
[\mathrm{X}_{s}, \mathcal{F}^{sa}_{*}(\mathscr{H})] &\approx &[\mathrm{X}_{s}, \hat{\mathcal{F}}_{*}(\mathscr{H})]\,, \nonumber\\
&\approx & [\mathrm{X}_{s}, \Omega\mathcal{F}(\mathscr{H})]\,, \nonumber\\
&\approx & [\mathit{S}\mathrm{X}_{s}\,, \mathcal{F}(\mathscr{H})]_{*}\,,\nonumber\\
&\approx & \mathit{K}^{-1}(\mathrm{X}_{s})\,,
\end{eqnarray}
where $\mathit{K}^{-1}(\mathrm{X}_{s}) \approx \mathit{\tilde{K}}(\mathit{S}\mathrm{X}_{s,+})$ denotes the reduced $\mathit{K}$-group and $\mathit{S}\mathrm{X}_{s,+}$ is the suspension of our surface Brillouin zone with a disjoint point attached \cite{Hatcher-K-theory}. Notice that Grothendieck's completion popped out naturally from our construction!

\section{$\mathit{K}$-theory or vector bundles?}
Geometric constructions of fermionic topological phases on an infinite crystal with no boundary often lead to isomorphism classes of $n$-dimensional vector bundles over the $d$-dimensional Brillouin zone $\mathrm{X}$, $\mathit{Vect}^{n}_{\mathbb{C}}(\mathrm{X})$. In general, passing from $\mathit{Vect}^{n}_{\mathbb{C}}(\mathrm{X})$ to $\mathit{K}$-theory involves a series of steps. First we must allow addition of isomorphism classes $[E_{1}]\oplus[E_{2}] = [E_{1}\oplus E_{2}]$ and adding $1$-dimensional vector bundles (i.e. single bands), which forces us to consider isomorphism classes of vector bundles of any dimension over $\mathrm{X}$, arriving at the semigroup $\mathit{Vect}_{\mathbb{C}}(\mathrm{X})$. Then, we can either impose the stabilization equivalence relation \cite{Hatcher-K-theory}, as we did in subsection \ref{Horava-Local} to obtain the reduced $\tilde{\mathit{K}}(\mathrm{X})$ group or we can define another equivalence relation, defined on pairs $([E_1],[E_2]), ([E'_{1}],[E'_{2}])$ of isomorphism classes, turning $\mathit{Vect}_{\mathbb{C}}(\mathrm{X})$ into the abelian group $\mathit{K}(\mathrm{X})$ \cite{Hatcher-K-theory}. The latter process is called a Grothendieck completion or Grothendieck construction and is usually ignored in the physics literature, where most works compute the unreduced $\mathit{K}(\mathrm{X})$ group directly to classify fermionic gapped phases, without justifying the transition to $\mathit{K}$-theory. As adequately put by Thiang \cite{Thiang1}, this has produced a \textit{conflation} between the full $\mathit{K}(\mathrm{X})$ groups and their reduced versions $\tilde{\mathit{K}}(\mathrm{X})$, leading to discrepancies among some tables of gapped topological phases \cite{Thiang1}.

The mathematical literature on the other hand addresses the issue in different ways. As examples we can consider the work of De Nittis and Gomi \cite{D-G-AI},\cite{D-G-AII}, which uses $n$-dimensional vector bundles and their characteristic classes to classify fermionic phases, without appealing to $\mathit{K}$-theory. In a different vein Freed and Moore \cite{Freed-Moore} employ a process equivalent to a Grothendieck completion but emphasize that they are not aware of an apriori good physical motivation for its use. There are also examples in the $C^{*}$-algebraic approach of Prodan and Schulz-Baldes \cite{Prodan-Schulz-Baldes} together with references therein. Prodan and Schulz-Baldes motivate the use of $\mathit{K}$-theory for lattice Hamiltonians in the so called \textit{tight-binding} approximation \cite{Ashcroft-Mermin}, through its applications in the bulk-boundary correspondence when combined with \textit{cyclic cohomology}. Though a classification using $\mathit{K}$-theory is not a direct logical consequence of their formalism, its use in their approach is certainly well motivated and it employs $\mathit{K}$-theory of $C^{*}$-algebras for both surface states and bulk states, with similar results as ours for the surface states in symmetry class $A$, employing the functor $\mathit{K}^{-1}$ but on a \textit{real space} algebra of surface observables.

To our knowledge, our construction for systems on a half-space is the only one for which $\mathit{K}$-theory is not put in by hand but rather arises naturally from the simple mathematical interpretations of adiabatic evolution and a gapped bulk spectrum for Bloch Hamiltonians on a half-space. We also do not restrict to Hamiltonians in the tight-binding approximation nor lattice systems, nevertheless our construction has the rather strong assumption of discrete translation-invariance on the boundary and the more common assumption of a gapped bulk spectrum, restrictions which are better handled in \cite{Prodan-Schulz-Baldes}.

 \section{Adding symmetries}
 \label{Sec:Symmetries}
 Let us mention from the offset that we will only study systems on a half-space for which the boundary conditions preserve the symmetry class of the corresponding infinite crystal (bulk) Hamiltonian. Thus, both the continuous and discrete spectrum of our Hamiltonians $\sigma_{d}(\mathcal{H}(\vec{k}_{\parallel}))$, $\sigma_{c}(\mathcal{H}(\vec{k}_{\parallel}))$ respectively, inherit the same structure when a symmetry is implemented.  
\subsection{Real and quaternionic structure on a complex vector space} 
  We must first digress and introduce the following mathematical definitions taken from \cite{Moore-Real}. A \textit{real} structure on a complex vector space $\mathrm{V}$ is an \textit{anti-linear} operator $K$, such that $K^2 = I$.
In this case $\mathrm{V} \approx \mathrm{W}\otimes_{\mathbb{R}} \mathbb{C}$, where $\mathrm{W}$ is vector space over $\mathbb{R}$.
A \textit{quaternionic} structure on a complex vector space $\mathrm{V}$ is an \textit{anti-linear} operator $K$, such that $K^2 = -I$. $I, K, i$ and $J= iK$ satisfy the quaternion relations. As will be shown below these structures will be induced on our Hilbert space by the corresponding symmetry operators and will modify the complex $\mathit{K}$-theory that appears in our classification into a different (but similar) extraordinary cohomology theory \cite{Hatcher-Alg-Top}, that we will use to compute the distinct classes of globally topologically stable Fermi surfaces in a given symmetry class.

\subsection{Particle-hole symmetry}
The first symmetry we consider is the so called \textit{particle-hole} symmetry. Systems with only particle-hole symmetry and discrete translation-invariance are denoted as symmetry classes $C$ and $D$ in the Altland-Zirnbauer classification \cite{Altland-Zirnbauer} and the operator representing said symmetry is traditionally denoted by $\Xi$. There are different choices of implementations of $\Xi$ one can make, such as whether it is a unitary or antiunitary operator \cite{Freed-Moore},\cite{Thiang1} and each one of them represents systems in condensed matter with different physical properties. Our choice of implementation is the one most commonly used throughout the physics literature \cite{Kitaev},\cite{Ryu-Fermi} \cite{Hasan-Kane}, where our symmetry operator $\Xi$ satisfies
\begin{eqnarray}
\Xi \mathcal{H}(\vec{k}_{\parallel}) &=& -\mathcal{H}(-\vec{k}_{\parallel})\Xi\,,\label{eq:Real1}\\
\Xi i &=& -i\Xi\,.\label{eq:Real2}
\end{eqnarray}
This implies that for each spectrum index $\alpha$ of $\mathcal{H}(\vec{k}_{\parallel})$, there exists a spectrum index $\beta$ of $\mathcal{H}(-\vec{k}_{\parallel})$, such that  
 \begin{eqnarray}
 \Xi \ket{\phi_{\alpha}(\vec{k}_{\parallel},r_{\perp})} &=& \ket{\phi_{\beta}(-\vec{k}_{\parallel},r_{\perp})}\,,\\
 \varepsilon_{\alpha}(\vec{k}_{\parallel}) &=&-\varepsilon_{\beta}(-\vec{k}_{\parallel})\,.\label{PH-spectrum} 
 \end{eqnarray}
 In general $\beta \neq \alpha$ to respect the symmetry, however if $\varepsilon_{\alpha}(\vec{k}_{\parallel})\in \sigma_{d}(\mathcal{H}(\vec{k}_{\parallel}))$ passes through an involution fixed-point $\vec{k}_{\parallel}= -\vec{k}_{\parallel}$, then $\beta$ may equal $\alpha$. This is a subtle but non-trivial point as it permits the existence of some topologically non-trivial Fermi surfaces for all symmetry classes considered here (see subfigures \ref{fig:d=2-ClassC}, \ref{fig:d=2-ClassD}). Note that for particle-hole symmetric systems there are no $\mathcal{F}^{sa}_{\pm}$-components due to equation (\ref{PH-spectrum}). 
\subsubsection{Class $D$} 
If our system belongs to Class $D$, we must further assume that
\begin{equation}
\Xi ^2 = I \label{eq:Real3}\,.
\end{equation} 
  Thus, $\Xi$ represents a \textit{real} structure on our complex Hilbert space $\mathscr{H}$, as defined above, which we denote as $\mathscr{H}_{R}$. It also induces an \textit{involution} $\tau : \mathrm{X}_{s} \rightarrow \mathrm{X}_{s}$, $\tau (\vec{k}_{\parallel}) = -\vec{k}_{\parallel}$, making $\mathrm{X}_{s}$ a \textit{real} space in the sense of \cite{Atiyah-Real}. The general strategy for extending the results of \cite{Atiyah-Skew} is to employ \textit{equivariant} homotopy \cite{May-Equiv}, where the above implementation of the symmetry induces the action $\mathcal{H}(\vec{k}_{\parallel}) \mapsto -\Xi \mathcal{H}(-\vec{k}_{\parallel})\Xi$. We wish to use a slightly different action, defined in \cite{Matumoto}, for which we must pass to the skew-adjoint operator $i\mathcal{H}(\vec{k}_{\parallel})$, with the action 
 \begin{equation}
 \label{eq:action1}
  i\mathcal{H}(\vec{k}_{\parallel}) \mapsto \Xi i \mathcal{H}(-\vec{k}_{\parallel})\Xi.
 \end{equation}
This subtle difference allows us to employ a result of Matumoto \cite{Matumoto}
 \begin{equation}
 \label{eq:Matumoto}
 [\mathrm{X}_{s}, \mathcal{F}(\mathscr{H}_R)]_{C_2} \approx \mathit{KR}(\mathrm{X}_{s})\,.
 \end{equation} 
 where $\mathit{KR}$ is a different kind of $\mathit{K}$-theory, originally developed in \cite{Atiyah-Real} and $C_{2}$ denotes the cyclic group of order $2$, with the associated action (\ref{eq:action1}). Using Whitehead's equivalence in conjunction with (\ref{eq:Matumoto}), it suffices to have plain homotopy equivalence on the fixed-points of the subgroups of $C_2$ to have the full equivariant equivalence, yielding 
 \begin{equation}
 [\mathrm{X}_{s}, \hat{\mathcal{F}}(\mathscr{H}_R)]_{C_2} \approx \mathit{KR}^{-1}(\mathrm{X}_{s})\,.
 \end{equation} 
The equivalence on the corresponding fixed-points is given by the Atiyah-Singer map employed in \cite{Atiyah-Skew}, said map being equivariant under the chosen action. This group represents the topologically stable isomorphism classes of Fermi surfaces in symmetry class $D$. After seeing the implications particle-hole symmetry has for $\sigma(\mathcal{H}(\vec{k}_{\parallel}))$ it becomes clear that such a symmetry cannot be fully realized in condensed matter systems, nevertheless it is a powerful approximation to employ in the analysis of superconductors \cite{Superconductivity} where it is true for states near the superconducting gap $\Delta$.
\subsubsection{Class $C$}
Now we move on to the case where 
\begin{equation}
\Xi^2 = -I\,
\end{equation}
 also known as symmetry class $C$. For this case, the particle-hole symmetry operator induces a \textit{quaternionic} structure on $\mathscr{H}$, $i\Xi$ being the fourth generator. Let us denote this quaternionic structure by $\mathscr{H}_Q$. Again we employ the same strategy as above, passing to skew-adjoint operators where, following \cite{Matumoto} we denote by $D_2$ the cyclic group of order $2$, which acts on $\mathcal{F}(\mathscr{H}_Q)$ by the different action
 \begin{equation}
 \label{eq:action2}
  i\mathcal{H}(\vec{k}_{\parallel}) \mapsto -\Xi i \mathcal{H}(-\vec{k}_{\parallel})\Xi\,.
 \end{equation}
 Matumoto also proved the equivalence
  \begin{equation}
 \label{eq:Matumoto2}
 [\mathrm{X}_{s}, \mathcal{F}(\mathscr{H}_Q)]_{D_2} \approx \mathit{KH}(\mathrm{X}_{s})\,.
 \end{equation}
$\mathit{KH}$ is known as \textit{symplectic} $\mathit{KR}$-theory, and we will follow the notation employed in \cite{Rosenberg-Elliptic}.	Repeating the fixed-point argument as above, we conclude that stable isomorphism classes of Fermi surfaces in class $C$ are given by
\begin{equation}
 [\mathrm{X}_{s}, \hat{\mathcal{F}}(\mathscr{H}_Q)]_{D_2} \approx \mathit{KH}^{-1}(\mathrm{X}_{s})\,.
\end{equation}
 We can compute $\mathit{KH}^{-n}(\mathrm{X})$ for any $n$ entirely in terms of $\mathit{KR}$-groups, since we have the isomorphism
\begin{equation}
\label{eq:KHtoKR}
\mathit{KH}^{-n}(\mathrm{X}) \approx \mathit{KR}^{-n-4}(\mathrm{X})\,.
\end{equation}
\subsection{Time-reversal symmetry}
We shall implement \textit{time-reversal symmetry} through an operator $\Theta$ \cite{Kitaev},\cite{Hasan-Kane}
\begin{eqnarray}
\Theta \mathcal{H}(\vec{k}_{\parallel}) &=& \mathcal{H}(-\vec{k}_{\parallel})\Theta\,,\label{eq:Real4}\\
\Theta i &=& -i\Theta\,.\label{eq:Real5}
\end{eqnarray}
Similarly to particle-hole symmetry, this implies that for each spectrum index $\alpha$ of $\mathcal{H}(\vec{k}_{\parallel})$, there exists a spectrum index $\beta$ of $\mathcal{H}(-\vec{k}_{\parallel})$, such that  
 \begin{eqnarray}
 \Theta\ket{\phi_{\alpha}(\vec{k}_{\parallel},r_{\perp})} &=& \ket{\phi_{\beta}(-\vec{k}_{\parallel},r_{\perp})}\,,\\
 \varepsilon_{\alpha}(\vec{k}_{\parallel}) &=& \varepsilon_{\beta}(-\vec{k}_{\parallel})\,.\label{TR-spectrum} 
 \end{eqnarray}

There are two classes in the Altland-Zirnbauer classification, that correspond to systems with only discrete translation-invariance and time-reversal symmetry, namely classes $AI$ and $AII$, whose properties we describe below.
\subsubsection{Class $AI$}
Systems in class $AI$ must obey
\begin{equation}
\Theta ^2 = I\label{eq:Real6}\,.
\end{equation}
We now have the appropriate action directly on self-adjoint operators since 
\begin{equation}
\label{eq:action3}
\mathcal{H}(\vec{k}_{\parallel}) \mapsto \Theta \mathcal{H}(-\vec{k}_{\parallel})\Theta\,.
\end{equation}
 Looking again at the fixed-points of the subgroups of $C_2$ using the homotopy equivalence 
 \begin{equation}
 \mathcal{F}^{sa}_{*}(\mathscr{H}) \simeq \Omega ^{7}\mathcal{F}(\mathscr{H}^{'})
 \end{equation}
proven in \cite{Atiyah-Skew}, which holds for the fixed-points of all subgroups of $C_2$ and where $\mathscr{H}^{'}$ is a $*$-represantation of the $\mathit{C}_{6}$ Clifford algebra or its complexification, depending on the choice of subgroup \cite{Atiyah-Skew}. This equivalence is equivariant with respect to (\ref{eq:action1}), hence, together with (\ref{eq:Matumoto}) this implies that
\begin{equation}
 [\mathrm{X}_{s}, \mathcal{F}^{sa}_{*}(\mathscr{H}_R)]_{C_2} \approx \mathit{KR}^{-7}(\mathrm{X}_{s})\,,
 \end{equation}
 corresponds to isomorphism classes of stable Fermi surfaces in symmetry class $AI$.
 \subsubsection{Class $AII$}
Finally, for stable Fermi surfaces in symmetry class $AII$, we must assume
\begin{equation}
\Theta^2 = - I\,.
\end{equation}
By repeating the same process for self-adjoint operators as we did for skew-adjoint operators in class $C$, since
\begin{equation}
\label{eq:action4}
\mathcal{H}(\vec{k}_{\parallel}) \mapsto -\Theta \mathcal{H}(-\vec{k}_{\parallel})\Theta
\end{equation}
and we obtain
\begin{equation}
 [\mathrm{X}_{s}, \mathcal{F}^{sa}_{*}(\mathscr{H}_Q)]_{D_2} \approx \mathit{KH}^{-7}(\mathrm{X}_{s})\,.
 \end{equation}
 We remark that both $\Xi$ and $\Theta$ are either real or quaternionic structures on our physical Hilbert space and that the difference between them lies in the set of fixed-points under (\ref{eq:action1}), (\ref{eq:action2}), (\ref{eq:action3}), (\ref{eq:action4}) in the space of Hamiltonian operators $\mathcal{F}^{sa}_{*}(\mathscr{H}_R)$, $\mathcal{F}^{sa}_{*}(\mathscr{H}_Q)$ or their action-induced skew-adjoint counterparts. If we repeat the above process for classes $AI$ and $AII$ but instead look at equivariant homotopy classes of the $\mathcal{F}^{sa}_{+}$-component, corresponding to systems on an infinite crystal with no boundary, we see that they are \textit{equivariantly contractible}, proving our first stated conclusion in subsection \ref{Sub:IC}.
 
 \subsection{Relation to the work of Freed-Moore}
 A much more general method was developed by Freed and Moore \cite{Freed-Moore}, who undertook the arduous task of laying out all possible representations of symmetries and how they modify the classification for \textit{gapped} systems (for a $C^*$-algebraic version see \cite{Thiang1}). Unfortunately this method doesn't admit a trivial extension to our construction. In a few words, this happens because for the general case, symmetries need the compact-open topology but the space of Fredholm operators is contractible in this topology. 
The pattern underlying these examples and the work in \cite{Freed-Moore},\cite{Thiang1} suggests that to include symmetries it is necessary to have a generalized \textit{twisted equivariant} $\mathit{K}$-theory \cite{Adem-Twisted},\cite{Karoubi-Twisted}, that reduces to the variants we have mentioned here.
\section{Analysis}
We shall compute different classes of topologically stable Fermi surfaces for the five cases discussed above (discrete translation-invariance, particle-hole symmetry with $\Xi^2 = \pm I$ and time-reversal symmetry with $\Theta^2 = \pm I$ ) for $d$-dimensional systems on a half-space, i.e. when the surface Brillouin zone $\mathrm{X}_{s} = \mathbb{T}^{d-1}$, a $d-1$-dimensional torus or its involutive version $\bar{\mathbb{T}}^{d-1}$ for $d = 1,2,3$. We only present with some detail the computations of the different classes of Fermi surfaces for the only non-trivial case when $d =1$-dimensional systems on a half-line and $d=2$-dimensional systems on a half-plane (Fermi points).
\subsection{$d=1$ systems}
\begin{figure}[]
%\captionsetup[subfigure]{labelformat=parens}
\captionsetup{justification=raggedright}
  	\centering
  	%\vspace{1cm}
  \subfloat[ ]{\label{fig:d=1-Trivial}\includegraphics[width=0.08\textwidth]{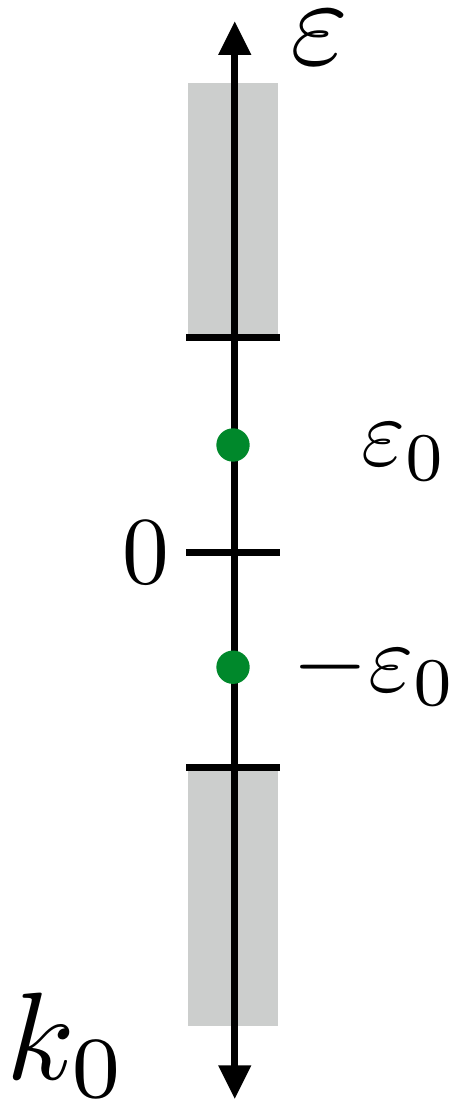}}
  \subfloat[ ]{\label{fig:d=1-Majorana}\includegraphics[width=0.1\textwidth]{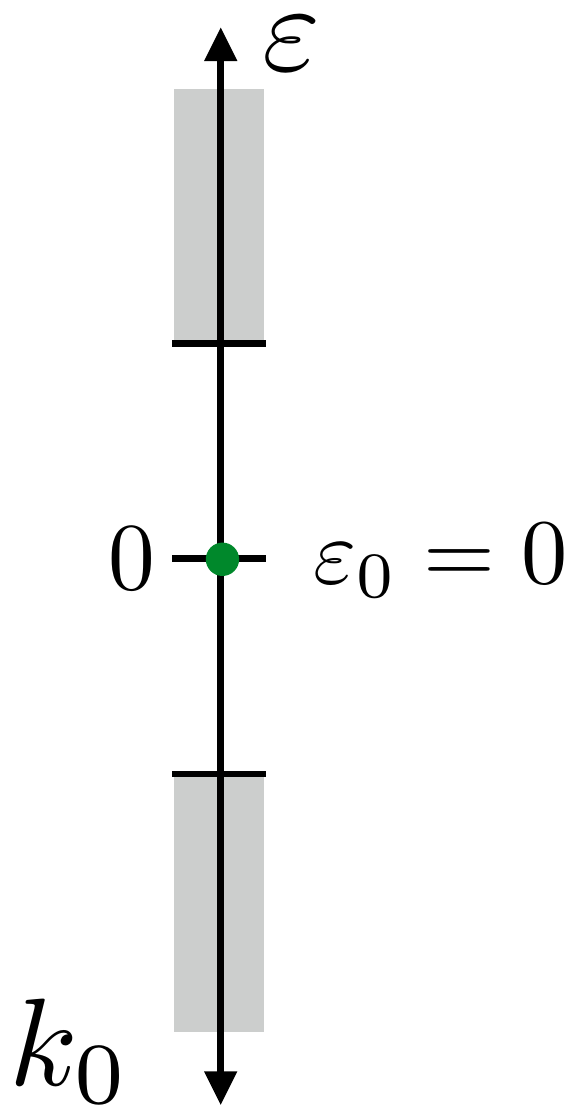}}
	 	 \caption[]{ 
$d=1$ unstable and stable modes. \ref{fig:d=1-Trivial} has no zero modes, where as \ref{fig:d=1-Majorana} represents a topologically stable Majorana zero mode as an edge state. $k_0$ is the $0$-dimensional surface Brillouin zone.} 
\end{figure}

The only non-trivial example for $d =1$, $\mathrm{X}_{s} = {k_0}$, a point, is in symmetry class $D$, where 
\begin{eqnarray}
\mathit{KR}^{-1}(k_0) &\approx & \tilde{\mathit{KR}}(\mathit{S}^{1})\,,\nonumber\\
&\approx & \tilde{\mathit{KO}}(\mathit{S}^{1})\,\nonumber\\
& \approx & \mathbb{Z}_{2}
\end{eqnarray}
The non-trivial element of $\mathbb{Z}_{2}$ represents the so called Majorana chain developed by Kitaev \cite{Kitaev-Majorana}, shown in subfigure \ref{fig:d=1-Majorana}. The single-particle picture corresponds to a \textit{Bogoliubov quasi-particle} and we have taken the \textit{mean field} approximation to electron-electron and electron-phonon interactions \cite{Superconductivity}. Note that the validity of adiabatic evolution (Fermi liquid theory) for $d=1$-systems is strongly limited.

\subsection{$d =2$ systems}
\begin{figure}[]
%	\captionsetup[subfigure]{labelformat=parens}
\captionsetup{justification=raggedright}
  	\centering
  	%\vspace{1cm}
	\subfloat[ ]{\label{fig:d=2-ClassA}\includegraphics[width=0.25\textwidth]{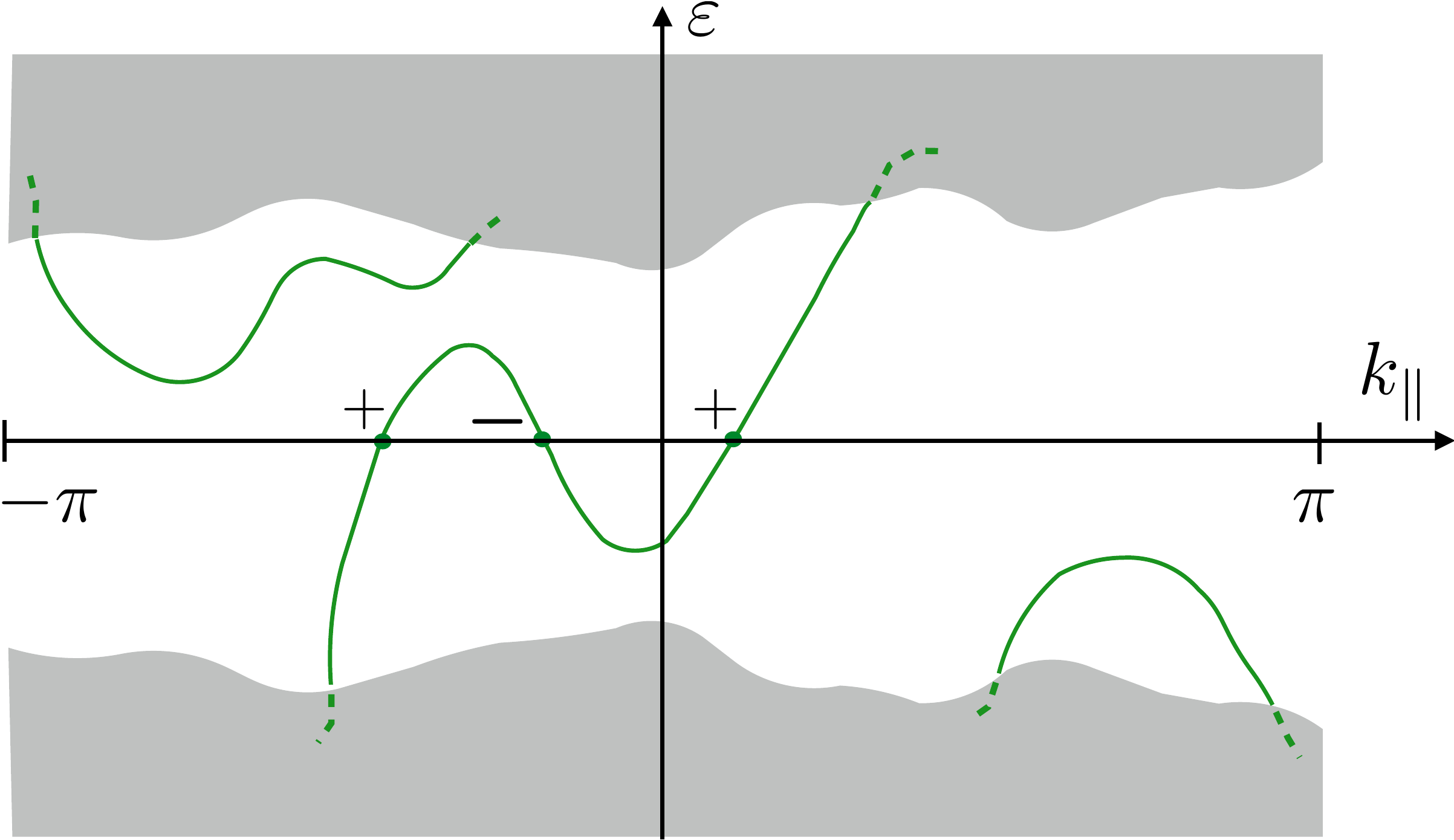}}
	  \subfloat[ ]{\label{fig:d=2-ClassC}\includegraphics[width=0.25\textwidth]{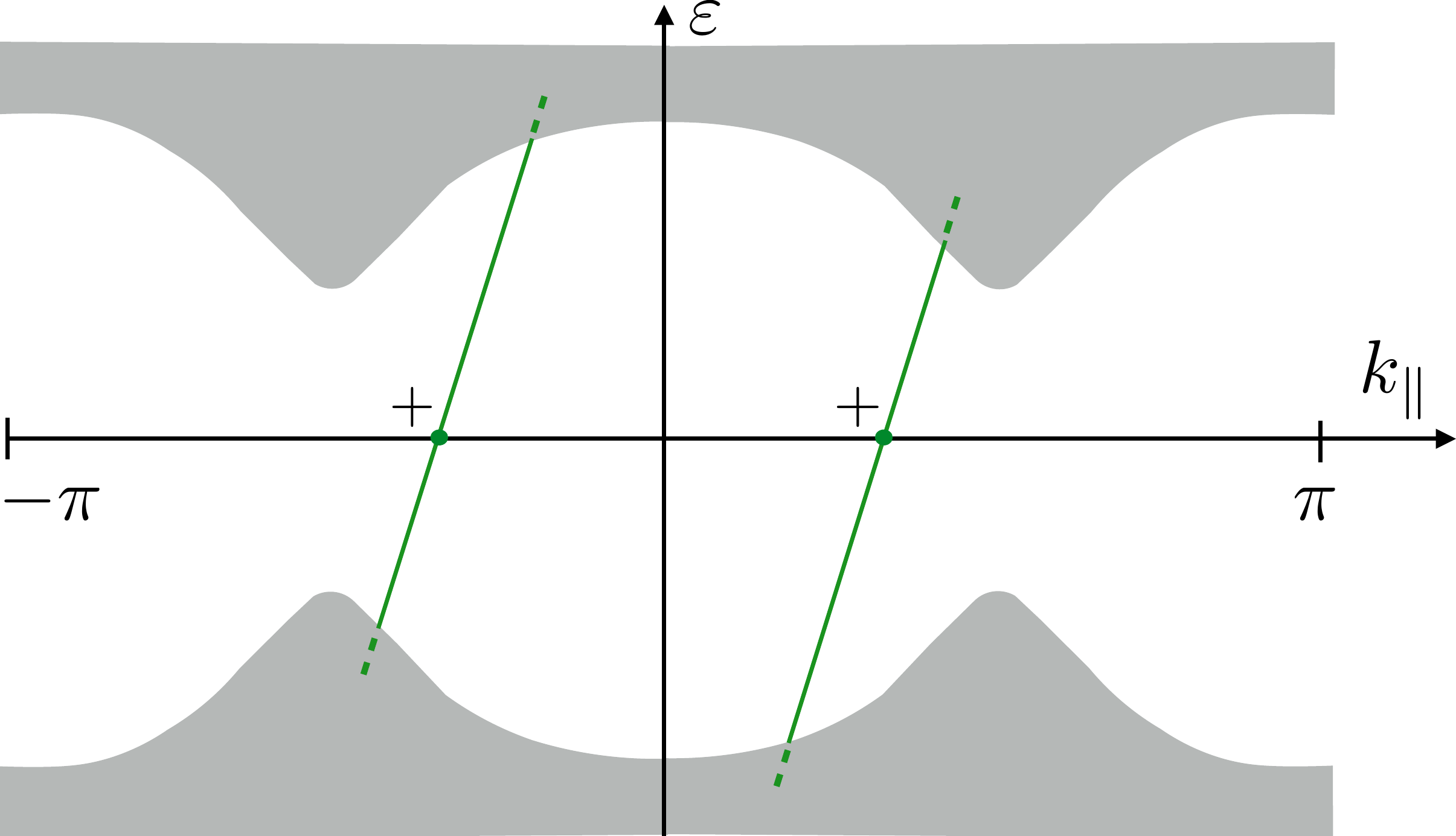}}\quad
\subfloat[ ]{\label{fig:d=2-ClassD}\includegraphics[width=0.25\textwidth]{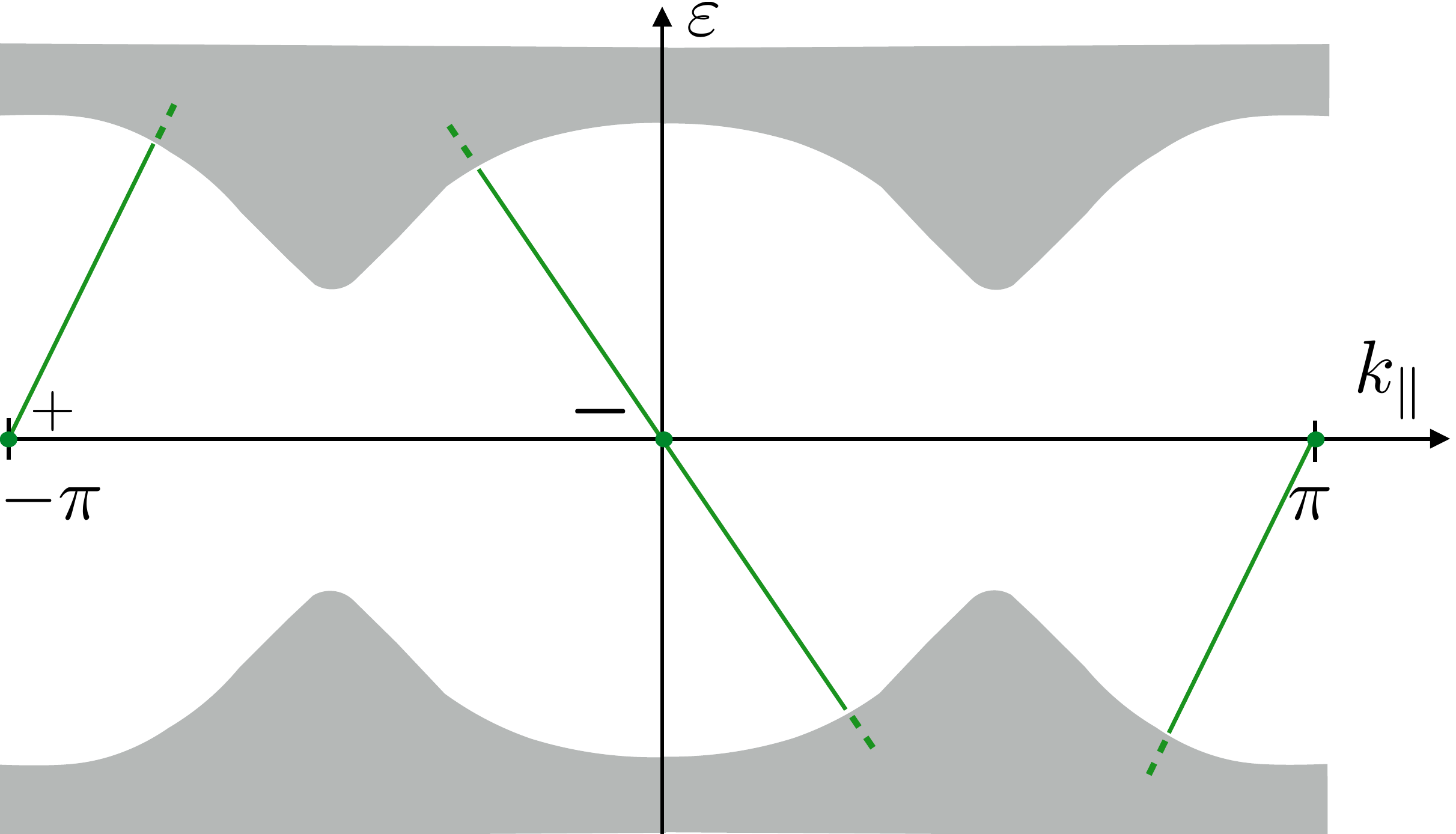}}
  \subfloat[ ]{\label{fig:d=2-ClassAII}\includegraphics[width=0.25\textwidth]{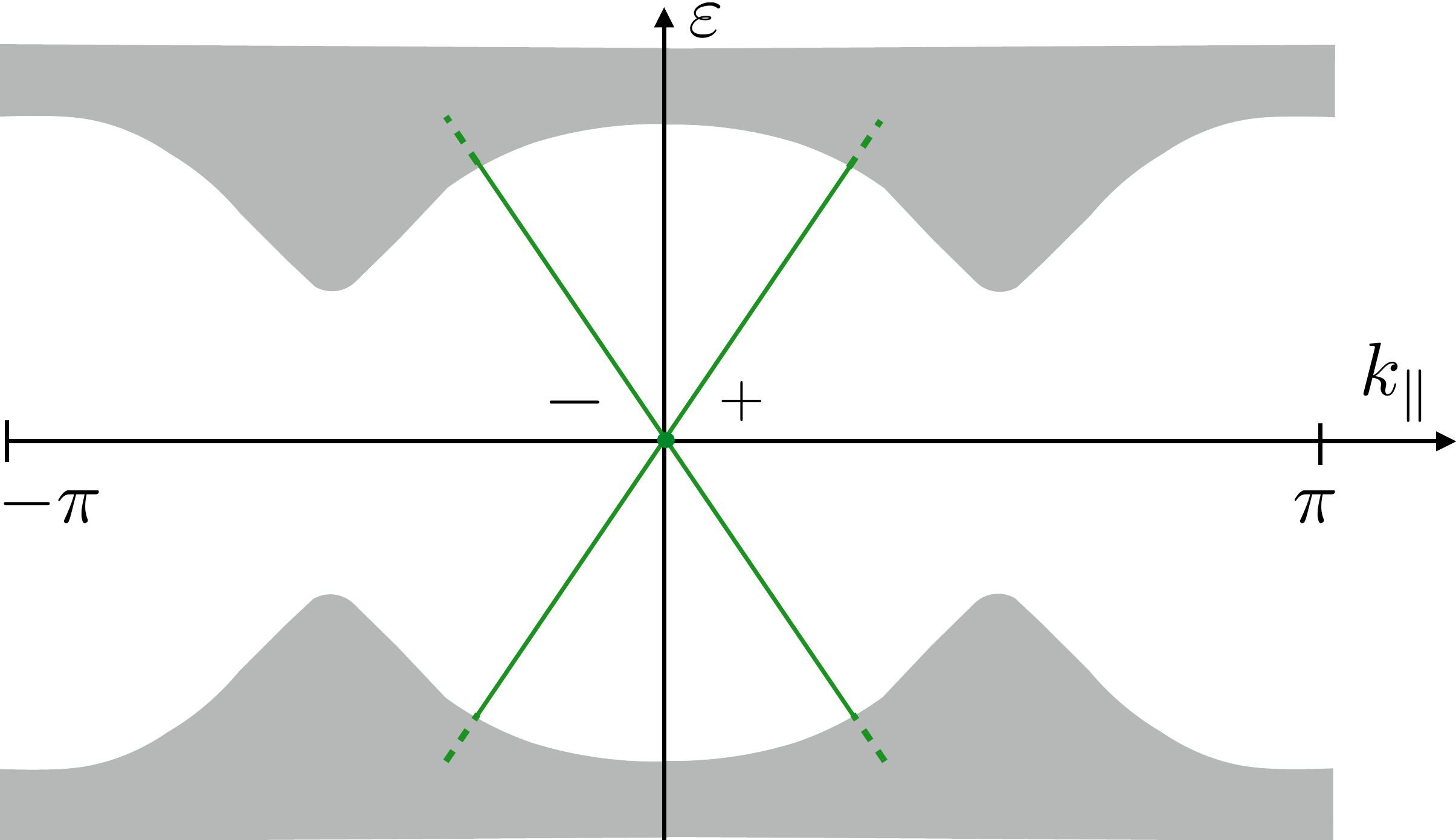}}

	\caption[]{ $d=2$ Topologically stable Fermi points. Solid regions and dashed lines belong to the continuous spectrum $\sigma_{c}(\mathcal{H}(\vec{k}_{\parallel}))$ (bulk states and resonance states respectively). Solid lines to $\sigma_{d}(\mathcal{H}(\vec{k}_{\parallel}))$ (surface state). $+$ ($-$) signs label the sign of the slope at crossings. \ref{fig:d=2-ClassA} corresponds to class $A$ (IQHE) with a net spectral flow = $1$. \ref{fig:d=2-ClassC} corresponds to a spectral flow of $2$, that may occur for both classes $C$ and $D$ at non fixed-points in  $\mathit{S}^{1,1}$. \ref{fig:d=2-ClassD} represents two Majorana zero modes in class $D$. \ref{fig:d=2-ClassAII} represents a helical Dirac fermion edge mode (SQHE).}
 \end{figure}

For $d=2$, our Brillouin zone $\mathrm{X}_{s} =  \mathit{S}^{1}$, is a circle. When we only have discrete translation-invariance,
\begin{eqnarray}
\mathit{K}^{-1}(\mathit{S}^1) &\approx & \tilde{\mathit{K}}(\mathit{S}^2)\,\nonumber\\
 &\approx & \mathbb{Z}\,.
\end{eqnarray}
The topological invariant that determines the corresponding class of a Fermi surface in $\mathit{K}^{-1}(\mathit{S}) \approx \pi_{1}(\mathcal{F}^{sa}_{*}(\mathscr{H}))$, is known as the \textit{spectral flow} \cite{Atiyah-Spectral}. This non-trivial state is shown in subfigure \ref{fig:d=2-ClassA}. The spectral flow had been employed previously to study chiral anomalies \cite{Semenoff-Spectral}. For this particular example, the physical interpretation of the spectral flow of our $2$-dimensional system corresponds to either bulk valence/conduction bands (positive/negative flow) having surface resonances, which become surface states, eventually cross the Fermi energy and become again surface resonance states in the bulk conduction/valence bands respectively. This amounts to a charge $U(1)$-chiral anomaly inflow \cite{Witten-FPI},\cite{Witten-Lectures} when an electric field is applied to the system. This is because an electron in a bulk valence/conduction band state becomes localized at the surface before eventually becoming a bulk conduction/valence state and it would appear that the boundary has gained/lost electric charge, when it is actually incoming/leaking from/into the bulk.

 Realizations of such systems are surface states of the IQHE \cite{Hasan-Kane}, once magnetic translations have been taking into account. Fukui and collaborators had already given a spectral flow interpretation to the IQHE \cite{Fukui-IQHE}, and further relate the spectral flow to a bulk \textit{Chern} number but did not relate it to $\mathit{K}$-theory as we have done in this work.

If we now consider a $2$-dimensional system with $\Xi^2 = I$, the involution $\tau(\vec{k}_{\parallel}) = -\vec{k}_{\parallel}$ is equivalent to having the Brillouin circle embedded in $\mathbb{C}$, with $\tau$ being complex conjugation. It has both $\{-1,1\}$ as its fixed-points and following \cite{Rosenberg-Elliptic} we denote our Brillouin circle with involution as $\mathit{S}^{1,1}$, where $\mathit{S}^{p,q}$ is the $(p+q-1)$-dimensional sphere embedded in $\mathbb{R}^{p,q} = \mathbb{R}^{p}\oplus i\mathbb{R}^{q}$. Thus we obtain
\begin{eqnarray}
\mathit{KR}^{-1}(\mathit{S}^{1,1}) &\approx & \mathit{KO}^{-1} \oplus \mathit{KO}\,\nonumber\\
 &\approx & \mathbb{Z}\oplus \mathbb{Z}_{2}\,,
\end{eqnarray}
where $\mathit{KO}^{j}$ denotes the real (trivial involution) $j$-th $\mathit{K}$-group of a point. For more details on how to compute $\mathit{KR}^{j}(\mathrm{X})$ we refer to \cite{Rosenberg-Elliptic}, section $3$. The $\mathbb{Z}$-factor again corresponds to a spectral flow of surface states similar to class $A$, but now $\forall\,\,\varepsilon_{\alpha}(\vec{k}_{\parallel}) \in \sigma(\mathcal{H}(\vec{k}_{\parallel}))$, there exists $\varepsilon_{\beta}(-\vec{k}_{\parallel}) \in \sigma(\mathcal{H}(-\vec{k}_{\parallel}))$ such that $\varepsilon_{\beta}(-\vec{k}_{\parallel}) = -\varepsilon_{\alpha}(\vec{k}_{\parallel})$. Thus for each zero mode at $\vec{k}_{\parallel}$, there must also be one at $-\vec{k}_{\parallel}$, as shown in subfigure \ref{fig:d=2-ClassC}. Hence the parity of the spectral flow in class $D$ depends on the existence of an odd number of zero modes at one of the involution fixed-points. 

The $\mathbb{Z}_{2}$-factor was discussed previously by Kitaev \cite{Kitaev} in the context of weak topological insulators. We first choose one of the involution fixed-points, then if there is an even/odd number of zero modes at the chosen point, we can determine if there is an even/odd number of zero modes at the other fixed-point by computing the parity of the spectral flow. In subfigure \ref{fig:d=2-ClassD} we represent a state with no net spectral flow, that has two zero modes, one at each fixed-point. This is the non-trivial $0\oplus 1$ state in $\mathbb{Z}\oplus\mathbb{Z}_{2}$. Let us remark that there is no canonical way of defining this invariant and a choice of a fixed-point must be taken. We also comment that the stability of this class depends on 1) respecting particle-hole symmetry and 2) the existence of 2 fixed-points on $\mathrm{X}_{s}$. In particular 2) does not hold for systems where there are \textit{umklapp} processes \cite{Ashcroft-Mermin}, leading to phenomena such as a Peierls's transition \cite{Peierls-Umklapp} on the boundary of our half-space. This makes the $\mathbb{Z}_2$-factor correspond to so-called \textit{weak} topological stability \cite{Weak-Topological}. We briefly discuss the connections of our work to weak topological phases in subsection \ref{subsection-Weak-Topological}. 

Examples of systems with a non-trivial class on the $\mathbb{Z}$-factor are spinless $p$-wave superconductors \cite{Hasan-Kane}, where Bogoliubov surface zero modes at one of the involution fixed-points $\{-1,1\}$, satisfy an analogous reality condition to that of \textit{Majorana} fermions, hence they are called \textit{Majorana zero modes}. Bogoliubov quasi-particles are effectively charge-less, thus, our zero modes only contribute to the thermal current. Such a current could be induced by a thermal gradient on our system, similar to the electric field for class $A$ \cite{Ryu-Thermal}. This spectral flow has also an anomaly inflow interpretation, where the anomaly is analogous to a perturbative gravitational anomaly \cite{Ryu-Thermal}.

For symmetry class $C$, using eq. (\ref{eq:KHtoKR}) we have
\begin{eqnarray}
\mathit{KH}^{-1}(\mathit{S}^{1,1}) &\approx & \mathit{KO}^{-5} \oplus \mathit{KO}^{-4}\,\nonumber\\
 &\approx & \mathbb{Z}\,.
\end{eqnarray}
 Since now $\Xi^2 = -I$, Fermi points at the involution fixed-points must be doubly degenerate and the $mod\, 2$ invariant must always vanish. This is similar to Kramer's theorem for time-reversal symmetry. Thus the spectral flow must come in pairs and there are no Majorana zero modes. There is the same perturbative gravitational anomaly inflow interpretation as in symmetry class $D$ (subfigure \ref{fig:d=2-ClassC}) for the Bogoliubov zero modes. Condensed matter systems with $\Xi^2 = -I$ in $d=2$ correspond to so-called $d$-wave superconductors \cite{Hasan-Kane}.

%\subsection{Higher dimensional cases}
%\begingroup
%\squeezetable
\begin{table}                 % optional [t, b or h];
\centering
\captionsetup{justification=raggedright}
  \begin{tabular}{ c | c  c | c  c c }
    \hline    
    \hline
    \multicolumn{3}{ c }{Symmetry } & \multicolumn{3}{ c }{$d$ }  \\
       $AZ$ &  \hspace{2em}$\Theta$ & \hspace{2em} $ \Xi$ &\hspace{2em}  $1$ &\hspace{2em}  $2$ &\hspace{2em}  $3$\\
    \hline
   $A$  & \hspace{2em}$ 0$ & \hspace{2em}$0$  & \hspace{2em}$0$ & \hspace{2em}$\mathbb{Z}$ & \hspace{2em}$\mathbb{Z}^2$\\
   $AI$  & \hspace{2em}$ I$ & \hspace{2em}$0$  & \hspace{2em}$0$ & \hspace{2em}$0$ & \hspace{2em}$0$\\
   $AII$  &\hspace{2em}$ -I$ & \hspace{2em}$0$  & \hspace{2em}$0$ & \hspace{2em}$\mathbb{Z}_{2}$ & \hspace{2em}$\mathbb{Z}_{2}^3$\\
   %\hline
    $D$ &\hspace{2em} $0$ & \hspace{2em}$I$ & \hspace{2em} $\mathbb{Z}_{2}$ & \hspace{2em}$\mathbb{Z}\oplus \mathbb{Z}_2$ & \hspace{2em}$\mathbb{Z}^2\oplus \mathbb{Z}_{2}$\\
   % \hline
    $C$  &\hspace{2em} $0$ & \hspace{2em}$-I$  & \hspace{2em} $0$ & \hspace{2em}$\mathbb{Z}$ & \hspace{2em}$\mathbb{Z}^2$\\
    
    \hline
    \hline
  \end{tabular}
  \caption[]{Classes of topologically stable Fermi surfaces for half-space systems with a gapped bulk in dimension $d = 1,2$ and $3$. $\Theta$ is the time-reversal operator and $\Xi$ is the particle-hole symmetry operator. We use the names given in \cite{Altland-Zirnbauer} ($AZ$). $0$ on the right-side denotes the trivial group. $0$ on the left-side indicates absence of said symmetry. When $\Theta = \Xi = 0$ the surface Brillouin zone $\mathrm{X}_{s}$ is a regular torus $\mathbb{T}^{d-1}$ and for either $\Theta^2 = \pm I$ or $\Xi^2 = \pm I$ it is a torus with involution $\bar{\mathbb{T}}^{d-1}$. Higher-dimensional classes relate to weak bulk topological phases.
    \label{tab:Results}}
  \end{table}
%\endgroup

 For symmetry classes $AI$ and $AII$ the spectral flow must vanish because of the symmetry of the spectrum $\varepsilon_{\beta}(-\vec{k}_{\parallel}) = \varepsilon_{\alpha}(\vec{k}_{\parallel})$ for some $\alpha,\beta$. By computing class $AI$, $d=2$
\begin{eqnarray}
\mathit{KR}^{-7}(\mathit{S}^{1,1}) &\approx & \mathit{KO}^{-7} \oplus \mathit{KO}^{-6}\,\nonumber\\
 &=& 0\,.
\end{eqnarray}  
We see there is only the trivial class of Fermi points, i.e. all Fermi points are adiabatically gappable in class $AI$, $d=2$. For symmetry class $AII$, $d =2$ we obtain
\begin{eqnarray}
\mathit{KH}^{-7}(\mathit{S}^{1,1}) &\approx & \mathit{KO}^{-3} \oplus \mathit{KO}^{-2}\,\nonumber\\
 &\approx &\mathbb{Z}_{2}\,.
\end{eqnarray} 
 Due to Kramer's theorem, if Fermi points arise at the involution fixed-points $\{1,-1\}$ of $\mathrm{X}_{s}$, they must be doubly degenerate, and thus the invariant is the number of double-crossed Fermi points $mod\, 2$ \cite{Hasan-Kane}. Such double-crossed Fermi points have a $U(1)$-\textit{spin} anomaly inflow interpretation \cite{Ryu-Zhang} and are often called helical Dirac zero modes because of the analogous role to the helicity of relativistic Dirac fermions played by spin-orbit coupling. A physical example is the quantum spin Hall effect (QSHE) \cite{Hasan-Kane}.
 
 \subsection{Higher dimensional cases and weak topological phases}
 \label{subsection-Weak-Topological}
   Our results for all dimensions and respective symmetry classes are summarized in Table \ref{tab:Results}. We only point to \cite{Rosenberg-Elliptic}, section $3$ and \cite{Freed-Moore}, Theorem $11.8$ where the methods for computing these groups are presented with the necessary amount of detail. For dimensions $2$ and $3$ Table \ref{tab:Results} shows non-trivial classes. One could worry that, using the bulk-boundary correspondence, these classes should appear in \cite{Hasan-Kane} and they do not. We note that we have assumed discrete translation-invariance, thus the proper correspondence would be between Table \ref{tab:Results} and weak topological phases \cite{Kitaev},\cite{Hasan-Kane}, that is, those which depend on discrete translation-invariance. These are usually constructed by stacking $d$-dimensional systems on top of each other, forming a $d+1$-dimensional system and looking for gapless edge excitations \cite{Hasan-Kane},\cite{Weak-Topological}. We also note that computions of weak topological phases sometimes yield more non-trivial classes than those which appear in Table \ref{tab:Results}. We believe this is due to the fact that our half-spaces have less boundaries than the prototypical construction employed to compute these, nonetheless a more detailed examination is needed to understand the connection between the two. We shall postpone the physical interpretation, possibly of anomaly inflow, of the higher dimensional cases for future work.
 \section{Conclusions}
We have rigorously derived a classification of topologically stable Fermi surfaces from electronic band theory and adiabatic evolution for systems on an infinite crystal and systems on a half-space. We rigorously proved that there can only be globally topologically unstable Fermi surfaces for all symmetry classes considered for systems on an infinite crystal. It remains to be seen to which, if not all, of the symmetry classes this result extends to.

We also adapted a simplified version of Ho\v{r}ava's local stability analysis of Fermi surfaces to the single-particle formalism and explained its relation to our global stability analysis. For systems on a half-space, the construction presented above yields the $\mathit{K}$-groups $\mathit{K}^{-1}(\mathrm{X}_{s}),\mathit{KR}^{-1}(\mathrm{X}_{s})$, $\mathit{KH}^{-1}(\mathrm{X}_{s}),\mathit{KR}^{-7}(\mathrm{X}_{s})$ and $\mathit{KH}^{-7}(\mathrm{X}_{s})$, with $\mathrm{X}_{s}$ the surface Brillouin zone, for the cases of discrete translation-invariance and the two different implementations of both particle-hole and time-reversal symmetry (classes $A,\,D,\,C,\,AI$ and $AII$). We also computed all possible classes for $\mathrm{X}_{s} = \mathbb{T}^{d-1},\,\bar{\mathbb{T}}^{d-1}$ and $d=1,2,3$, summarized in Table \ref{tab:Results}. We found that Fermi surfaces belonging to a non-trivial class have an anomaly inflow interpretation, with corresponding anomaly charge/spin chiral and gravitational anomalies \cite{Ryu-Zhang},\cite{Witten-FPI},\cite{Ryu-Thermal}. The nature of higher dimensional classes presented in Table \ref{tab:Results} seems to correspond to weak topological phases, but a more detailed analysis is needed.

 We mention in passing that our results for half-space systems strengthen the connection found by Ho\v{r}ava \cite{Horava} between Fermi surfaces and so-called \textit{D-branes} in Type $IIA$ and Type $I'A$ string theory \cite{Horava-Branes}, where Ho\v{r}ava obtained the same $\mathit{K}$-groups but of a compact space-time instead of the Brillouin zone.

 A crucial point clarified in this work is that the allowance of a continuous spectrum for the Bloch Hamiltonian $\mathcal{H}(\vec{k}_{\parallel})$ by the boundary of the half-space system permits a rigorous topological interpretation of non-trivial Fermi surfaces, without resorting to low-energy limits.
 
 Finally, it is necessary to develop a twisted formalism, so that we can include all possible symmetries as in \cite{Freed-Moore},\cite{Thiang1}, overcoming the apparent different choice between the norm and the compact-open topology and, if possible, to extend it so that in a single mathematical framework, we can encompass all topological phenomena of non-interacting fermions in condensed matter.

\appendix
\section{The Riesz metric for unbounded operators and Bloch's theorem}\label{Riesz-Topology}
We have employed the machinery of \cite{Atiyah-Skew},\cite{Atiyah-Fredholm} to derive a classification of globally stable Fermi surfaces and we have imposed that our Bloch Hamiltonian operator $\mathcal{H}(\vec{k})$ (or $\mathcal{H}(\vec{k}_{\parallel})$ respectively) be bounded. At first glance this imposition seems unjustified as equations (\ref{Infinite-Crystal-Bloch}), (\ref{Half-space-Bloch}) clearly represent unbounded operators. Nonetheless, let $\mathcal{CF}^{sa}(\mathscr{H})$ denote the set of not necessarily unbounded self-adjoint Fredholm operators. We define the Riesz metric on $\mathcal{CF}^{sa}(\mathscr{H})$ as
\begin{eqnarray}
L: \mathcal{CF}^{sa}(\mathscr{H})\rightarrow  \mathcal{F}^{sa}(\mathscr{H})\,,\\
\mathcal{H} \mapsto \mathcal{H} (I +\mathcal{H}^{\dagger}\mathcal{H})^{-\frac{1}{2}}\,,\\
d_{R}: \mathcal{CF}^{sa}(\mathscr{H}) \times \mathcal{CF}^{sa}(\mathscr{H}) \mapsto \mathbb{R}\,,\\
d_{R}(\mathcal{H}_1,\mathcal{H}_{2}) \mapsto \Vert L(\mathcal{H}_{1})-L(\mathcal{H}_2)\Vert ,
\end{eqnarray}
where $\Vert \centerdot \Vert$ denotes the standard operator norm and $L$ is known as the Riesz transform \cite{Lesch-Riesz}. The topology on $\mathcal{CF}^{sa}(\mathscr{H})$ induced by $d_{R}$ is known as the Riesz topology and the natural inclusion $\mathcal{F}^{sa}(\mathscr{H}) \hookrightarrow \mathcal{CF}^{sa}(\mathscr{H})$ is a homotopy equivalence. Hence $\mathcal{CF}^{sa}(\mathscr{H})$ splits into 3 path-components and $\mathcal{CF}^{sa}_{+}(\mathscr{H})$ is null-homotopic while $\mathcal{CF}^{sa}_{*}(\mathscr{H})$ is again a classifying space for the $\mathit{K}^{-1}$-functor \cite{Lesch-Riesz}. The issue with the choice of the Riesz topology is that it is generally hard to determine whether a family of operators in $\mathcal{CF}^{sa}(\mathscr{H})$ is continuous \cite{Lesch-Riesz}. However, as is shown in \cite{Feldman-Spectrum} for the infinite crystal, $\mathcal{H}(\vec{k})$ with $V(\vec{r})$ bounded has a fixed domain for all $\vec{k} \in \mathrm{X}$ and as explained in \cite{Boos-Riesz}, families of operators with a fixed domain are continuous in the Riesz topology. Similar arguments apply to the half-space Bloch Hamiltonian $\mathcal{H}(\vec{k}_{\parallel})$ as long as we fix the boundary conditions or, more generally, restrict to boundary conditions that have a fixed domain in $\mathscr{H}$. 

Given that
 \begin{eqnarray}
L(\Xi i\mathcal{H}) &=& \Xi L(i\mathcal{H})\,,\\
L(\Theta \mathcal{H}) &=& \Theta L(\mathcal{H})\,,
\end{eqnarray} 
 the equivariant results considered in section \ref{Sec:Symmetries} extend to the Riesz topology for $\mathcal{C}\hat{\mathcal{F}}(\mathscr{H}_{R}),$ $\mathcal{C}\hat{\mathcal{F}}(\mathscr{H}_{Q}),$ $\mathcal{CF}^{sa}_{*}(\mathscr{H}_{R})$ and $\mathcal{CF}^{sa}_{*}(\mathscr{H}_{Q})$ since the corresponding Bloch Hamiltonians and boundary conditions have a fixed domain in $\mathscr{H}_{R}$ or $\mathscr{H}_{Q}$, respectively.

\acknowledgments
D.Sheinbaum is in debt to A.K.Ramos-Musalem for preparing the figures in the text and to M.Fandino, J.Gonz\'alez-Anaya, B.Villareal and A.K.Ramos-Musalem for useful discussions. A.Adem and G.W.Semenoff were supported by research grants from NSERC, Canada. D.Sheinbaum is supported by a fellowship from CONACYT, Mexico (291928). 

\bibliographystyle{unsrt}
\small
\bibliography{paper-anom-JHEP-corrected}

\end{document}